\documentclass[aip, jcp, reprint, superscriptaddress, floatfix]{revtex4-2}

\pdfoutput=1

\usepackage{xcolor}
\usepackage{booktabs}
\usepackage{graphicx}
\usepackage{dcolumn}
\usepackage{amsmath}
\usepackage{xr-hyper}
\usepackage{hyperref}
\hypersetup{colorlinks=true, citecolor=blue, linkcolor=blue, urlcolor=blue}
\usepackage{stix}
\usepackage[Symbolsmallscale]{upgreek}

\newcommand{\black}{\textcolor{black}}

\newcommand{\emc}{e$\upmu$c}
\newcommand{\br}{\textbf{r}}
\newcommand{\bR}{\textbf{R}}
\newcommand{\erho}{\rho_\mathrm{e}}
\newcommand{\prho}{\rho_\mathrm{p}}
\newcommand{\arho}{\rho_\upalpha}
\newcommand{\brho}{\rho_\upbeta}
\newcommand{\mrho}{\rho_\upmu}
\newcolumntype{d}[1]{D{.}{.}{#1}}

\makeatletter
\newcommand*{\addFileDependency}[1]{
  \typeout{(#1)}
  \@addtofilelist{#1}
  \IfFileExists{#1}{}{\typeout{No file #1.}}
}
\makeatother
\newcommand*{\myexternaldocument}[2][]{%
    \externaldocument[#1]{#2}%
    \addFileDependency{#2.tex}%
    \addFileDependency{#2.aux}%
}
\myexternaldocument[supp:]{supplement}

\begin{document}

\title{Two-component density functional theory for muonic molecules: Inclusion of the electron-{positive} muon correlation functional}

\author{Mohammad Goli}
\email{m{\_}goli@ipm.ir}
\affiliation{School of Nano Science, Institute for Research in Fundamental Sciences (IPM), Tehran, 19395-5531, Iran}

\author{Shant Shahbazian}
\email{sh{\_}shahbazian@sbu.ac.ir}
\affiliation{Department of Physics, Shahid Beheshti University, Evin, Tehran, Iran}

\date{\today}

\begin{abstract}
It is well-known experimentally that the positively-charged muon and the muonium atom may bind to molecules and solids, and through muon’s magnetic interaction with unpaired electrons, valuable information on the local environment surrounding the muon is deduced. Theoretical understanding of the structure and properties of resulting muonic species requires accurate and efficient quantum mechanical computational methodologies. In this paper the two-component density functional theory, TC-DFT, as a first principles method, which treats electrons and the positive muon on an equal footing as quantum particles, {is} introduced and implemented computationally. The main ingredient of this theory, apart from the electronic exchange-correlation functional, is the electron-{positive} muon correlation functional which is foreign to the purely electronic DFT. A Wigner-type local electron-{positive} muon correlation functional, termed \emc-1, is proposed in this paper and its capability is demonstrated through its computational application to a benchmark set of {muonic} organic molecules. The TC-DFT equations containing \emc-1 are not only capable of predicting the muon’s binding site correctly but they also reproduce muon’s zero-point vibrational energies and the muonic densities much more accurately than the TC-DFT equations lacking \emc-1. Thus, this study set the stage for developing accurate electron-{positive} muon functionals, which can be used within the context of the TC-DFT to elucidate the intricate interaction of the positive muon with complex molecular systems.
\end{abstract}


\maketitle

\section{Introduction}%
The positively-charged muon, simply called muon hereafter, with a short half-life of $\sim$2.2 $\mu$s is one of those elementary particles that may bind and form bound states to atoms and molecules, termed muonic species, which are a subclass of the general class of the exotic species.\cite{nagamine_introductory_2003,horvath_exotic_2011} In fact, because of its similarity to proton, muon’s mass is $\sim$1/9th the proton’s mass,\cite{horvath_exotic_2011} muon is sometimes called the light radioisotope of hydrogen.\cite{rhodes_muoniumsecond_2002,rhodes_muoniumsecond_2012,macrae_isotopes_2013} If instead of muon, a muonium atom, Mu, composed of a muon and an electron, binds to a molecule the resulting muonic radical may react with other molecules opening a whole area called muonic chemistry.\cite{walker_muon_1983,walker_leptons_1985,roduner_positive_1988,roduner_polarized_1993,walker_kinetic_1998,mckenzie_using_2009,fleming_kinetic_2011,clayden_muons_2013,ghandi_muons_2015} These radicals are studied by a special spectroscopy called the muon spin resonance spectroscopy, $\upmu$SR,\cite{percival_theory_1976,heffner_muon_1984,brewer_prospects_1995,blundell_muon-spin_2004,brewer_problems_2012,mckenzie_positive_2013,yokoyama_future_2013,brewer_sr_2015} which may yield unique information on the binding site of muon in organic, organosilicon and biochemical molecule.\cite{nagamine_intra-_2000,mckenzie_reactions_2003,nagamine_electron_2004,macrae_structures_2006,jayasooriya_muon_2007,mccollum_probing_2008,mccollum_detection_2009,mccollum_reaction_2009,west_organosilicon_2010,mitra_silyl_2010,percival_free_2011,percival_dual_2012,mckenzie_muon_2014,west_germanium-centered_2014,west_silicon_2014,kobayashi_towards_2015,pant_hydration_2015,wright_muonium_2016,ito_observation_2018,mckenzie_radical_2018,chandrasena_free_2019,mckenzie_hydrogen-atom_2019,sugawara_opportunities_2019,samedov_free_2020,koshino_muonium_2021} Also, $\upmu$SR yields valuable information on details of muon’s binding site in condensed phases, for example local magnetic fields, which are hard to be deduced by other means.\cite{nagamine_introductory_2003,patterson_muonium_1988,cox_studies_1995,blundell_spin-polarized_1999,savici_muon_2002,menon_muons_2006,cox_muonium_2009,heron_muon-spin_2013,nuccio_muon_2014,yokoyama_photoexcited_2017,lord_optical_2020} In this regard, distinguishing the exact binding site of muon or Mu to complex molecules is vital for proper interpretation of $\upmu$SR spectrum although this is not an easy task since usually multiple sites, {that are} atoms and bonds, are available for binding. To overcome this problem theoretical modeling and computational considerations could be quite helpful and in the ideal situation, it is desirable to reproduce the whole $\upmu$SR spectrum computationally.\cite{bonfa_toward_2016} Our focus in this paper is also on new methodological developments in this direction. 

From a quantum mechanical viewpoint, muonic molecules may be treated in various ways since one may incorporate muon’s kinetic energy operator and its potential energy terms directly into the molecular Hamiltonian, conceiving it as a quantum particle, or treating it as a clamped particle like the heavy nuclei. The best approach is to treat all the constituents of a muonic molecule, {that are} electrons, muon and nuclei, equally as quantum particles where the path integral molecular dynamics is a vivid example.\cite{valladares_path-integral_1995,yamada_accurate_2014,oba_path_2016} However, the computational cost of such approaches is usually demanding, prohibiting their use for most molecules. Thus, the constituent particles are usually divided into subsets and each subset is considered separately within the field or the effective field of other particles. The most usual way of treating muonic molecules is to assume both muon and nuclei as clamped point charges at \black{the} first step and \black{treat} electrons as quantum particles. The widespread usage of this method is partly {due to the fact that} the {common} quantum computational packages may be employed directly for this purpose without further modifications. In this approach, based on the adiabatic approximation,\cite{mustroph_potential-energy_2016} the electrons of the molecule are first considered by the electronic Schr\"{o}dinger equation. Then, the muon and nuclei are assumed to be in the effective field produced by electrons and a Schr\"{o}dinger equation containing both muonic and nuclear variables is solved yielding basically the mixed muonic-nuclear vibrational normal modes. Hereafter we call this type of adiabatic approximation the single adiabatic electron/nuclei plus muon approximation or in short: SAA(e/$\upmu$n). Taking the fact that muon’s mass is somehow intermediate between electrons and nuclei, $m_\upmu/m_\mathrm{e} \approx 207$, $m_{\mathrm{proton}}/m_\upmu \approx 9$, $m_{^{12}\mathrm{C}}/m_\upmu \approx 106$ , the legitimacy of SAA(e/$\upmu$n) is not obvious particularly when compared to the mass ratio of even the lightest nuclei to electrons, for example $m_{\mathrm{proton}}/m_\mathrm{e} \approx 1836$. On the other hand, it is not also obvious whether muon’s vibrations are always coupled enough to heavier nuclei to justify simultaneous consideration of the nuclear and muonic vibrations. So, the mentioned Schr\"{o}dinger equation governing muonic and nuclear variables may in another adiabatic approximation be decoupled again. Accordingly, a single-particle Schr\"{o}dinger equation governs muon within the effective field produced by electrons and the clamped nuclei, and a Schr\"{o}dinger equation governs nuclei within the effective field produced by electrons and muon. In this approach there are three, instead of two, decoupled Schr\"{o}dinger equations governing muonic molecules, and thus it is called the double adiabatic approximation or in short DAA.\cite{porter_muonium_1999,bonfa_efficient_2015,wang_spintronic_2016} A third option for an adiabatic separation of particles in muonic molecules is to group electrons and the muon in a single packet while decoupling nuclei as a separate packet of particles. Hereafter we call this type of adiabatic approximation the single adiabatic electron plus muon/nuclei approximation or in short: SAA(e$\upmu$/n). To the best of authors' knowledge, this type of adiabatic separation has {rarely} been applied in computational studies,\cite{kerridge_quantum_2004,sheely_application_2010} and it is the premise of this study to find its justification and consider the full ramification of this approximation.           

The method of choice in computational consideration of the electronic structure of complex molecules and condensed phases is the electronic density functional theory (e-DFT).\cite{koch_chemists_2001,kohanoff_electronic_2006} While the e-DFT is intrinsically a single-component formalism, only assuming electrons as quantum particles, this is not an intrinsic limitation of the fundamental theorems of DFT,\cite{engel_density_2011} and the multi-component versions of DFT, MC-DFT, have {also} been formulated for systems composed of multiple types of quantum particles.\cite{sander_surface_1973,sander_surface_1973-1,kalia_surface_1978,dharma-wardana_density-functional_1982,capitani_nonbornoppenheimer_1982,nieminen_two-component_1985,boronski_electron-positron_1986,kryachko_formulation_1991,gidopoulos_kohn-sham_1998,kreibich_multicomponent_2001,van_leeuwen_first-principles_2004,leeuwen_erratum_2004,barnea_density_2007,kreibich_multicomponent_2008,messud_density_2009,messud_generalization_2011,gidopoulos_electronic_2014,kuriplach_improved_2014,wiktor_two-component_2015,kolesov_density_2018} Particularly, and relevant to present study, is the recent developments in formulation and computational applications of the MC-DFT to molecular systems where one or some numbers of protons, or other isotopes of hydrogen, are treated as quantum particles instead of clamped point charges.\cite{shigeta_density_1998,udagawa_hd_2006,pak_density_2007,imamura_colle-salvetti-type_2008,chakraborty_development_2008,chakraborty_properties_2009,imamura_extension_2009,chakraborty_erratum_2011,sirjoosingh_derivation_2011,sirjoosingh_multicomponent_2012,udagawa_electron-nucleus_2014,culpitt_multicomponent_2016,hashimoto_analysis_2016,udagawa_nuclear_2016,udagawa_unusual_2017,brorsen_multicomponent_2017,yang_development_2017,brorsen_alternative_2018,yang_multicomponent_2018,tao_multicomponent_2019,mejia-rodriguez_multicomponent_2019,xu_full-quantum_2020} In fact, since usually just two components of these systems are treated as quantum particles, for example electrons and protons, the general formulation of the MC-DFT is reduced to the two-component version, TC-DFT. The details of the TC-DFT formalism and its computational implementation will be considered in \black{the} next section but it suffices to mention that the main ingredients of the TC-DFT are the electronic exchange-correlation {functional} and the electron-proton correlation functional. The latter is a novel ingredient specific to the TC-DFT and is absent from the e-DFT because of the clamping of all nuclei. While in most practical applications of the TC-DFT the electron-proton correlation functional is usually completely neglected, this version may be called TC-DFT(ee),\cite{pak_density_2007} there {is} overwhelming evidence that this is not always justified and may bear serious errors {in} the prediction of certain properties of molecular systems.\cite{tao_multicomponent_2019,brorsen_multicomponent_2017} Particularly, the proton densities are over-localized in the absence of electron-proton functional and thus corresponding properties like zero-point energies, ZPEs, are {significantly} overestimated.\cite{tao_multicomponent_2019,brorsen_multicomponent_2017} Taking the fact that electron-nucleus correlation elevates by decreasing the mass of the nucleus,\cite{ito_formulation_2004} one expects that the electron-muon correlation in muonic molecules to be more important than the electron-proton correlation in protonic species. Thus, in this study our main goal is to introduce a simple but accurate electron-muon correlation functional for reliable modeling of muonic systems at the TC-DFT level. Let us stress that in our previous computational studies on muonic molecules,\cite{goli_deciphering_2014,goli_where_2015,goli_hidden_2015,goli_muon-substituted_2016,goli_how_2018} the electron-muon correlation was completely neglected and our recent formulation of the “effective electronic-only” DFT for muonic systems also does not contain this ingredient.\cite{rayka_effective_2018,goli_developing_2018}

The present paper is organized as follows. First, we will try to computationally justify the above proposed SAA(e$\upmu$/n) framework in a benchmark set of simple muonic organic molecules. Then, the theory and computational implementation of the TC-DFT for muonic molecules are discussed. The development and parameterization of our local electron-muon correlation functional, called \emc-1, is discussed subsequently and its properties are detailed. \black{Finally,} the resulting TC-DFT that contains both electronic exchange-correlation and electron-muon correlation functionals, TC-DFT(ee+e$\upmu$), is applied to the above-mentioned benchmark set and its significant superiority to the TC-DFT(ee) version is demonstrated.   

\section{Theory and computational implementation}

\subsection{Justifying the single adiabatic electron plus muon/nuclei approximation}

\begin{figure*}[t]
\includegraphics[width=\textwidth]{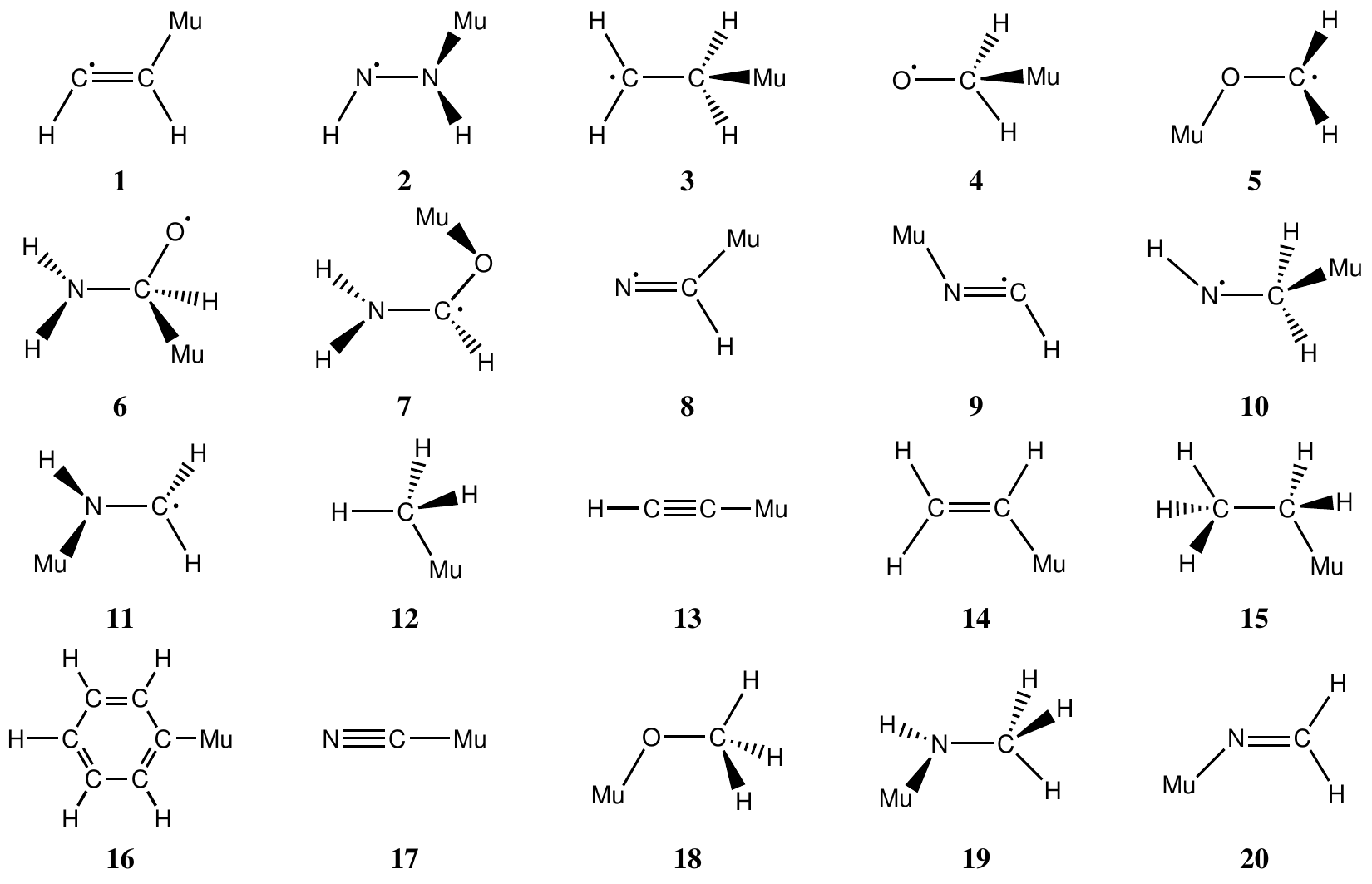}%
\caption{Schematic representations of the benchmark structures used in this study. The muonium atom is indicated by Mu symbol, and the single dot on atomic centers in \textbf{1}-\textbf{11} indicates the anticipated radical center generated after Mu addition.}
\label{fig:1}
\end{figure*}

\begin{figure*}[t]
\includegraphics[width=\textwidth]{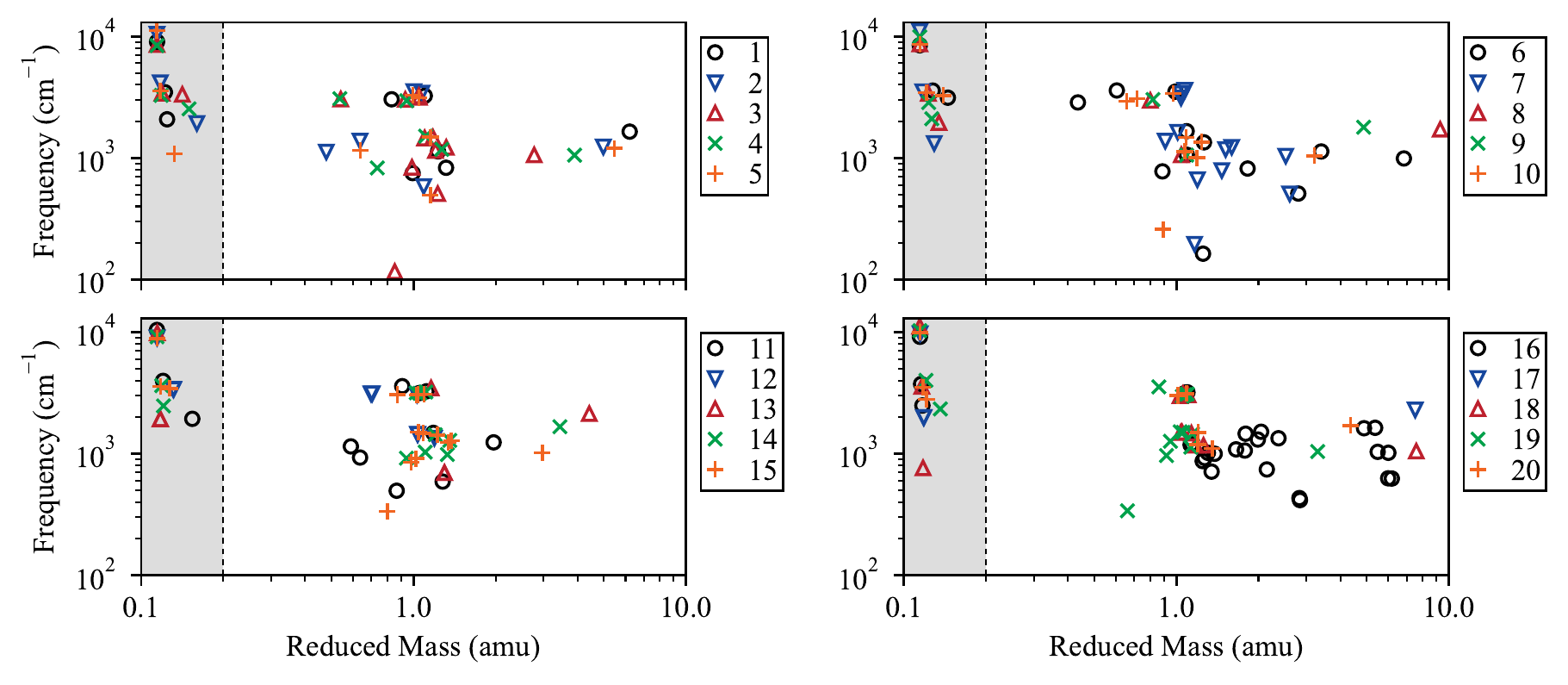}%
\caption{The distribution of frequencies versus reduced masses of the normal modes obtained based on the harmonic oscillator approximation at B3LYP/pc-2 level for \textbf{1}-\textbf{20}.}
\label{fig:2}
\end{figure*}

\begin{table}[t]
  \centering
  \caption{The muonic vibrational frequencies ($\mathrm{cm}^{-1}$) calculated within the harmonic oscillator approximation at B3LYP/pc-2 level and {from} 1D fittings to the 3D muon ground state wavefunction computed via the Numerov method. In case of B3LYP/pc-2, the on- and off-axis labels refer to the stretching and bending modes, respectively. In the case of the Numerov method, the on-axis vector refers to the direction parallel to the vector connecting the maximum of muon density and its binding atomic center, while the off-axis slice is perpendicular to the on-axis direction and also goes through the maximum of muon density.}
    \begin{ruledtabular}
    \begin{tabular}{l d{5.1}d{5.1}d{5.1}d{5.1}d{5.1}d{5.1}}
      & \multicolumn{3}{c}{B3LYP/pc-2} & \multicolumn{3}{c}{Numerov} \\
          \cmidrule(r{1pt}){2-4} \cmidrule(l{1pt}){5-7} 
     System & \multicolumn{1}{c}{On-axis}    & \multicolumn{1}{c}{Off-axis}   & \multicolumn{1}{c}{Off-axis}  & \multicolumn{1}{c}{On-axis}  & \multicolumn{1}{c}{Off-axis}   & \multicolumn{1}{c}{Off-axis}   \\ \hline
    1     & 9019.0 & 3469.9 & 2078.0 & 8466.6 & 3291.0 & 2111.0 \\
    2     & 10415.9 & 4157.0 & 1923.9 & 9751.1 & 3678.6 & 2142.1 \\
    3     & 8521.3 & 3401.8 & 3326.6 & 7838.0 & 3201.7 & 3034.9 \\
    4     & 8385.5 & 3267.3 & 2529.4 & 7738.3 & 3115.3 & 2746.7 \\
    5     & 11126.7 & 3545.8 & 1083.6 & 10463.9 & 3412.4 & 1590.6 \\
    6     & 8436.7 & 3576.4 & 3127.7 & 7770.6 & 3427.8 & 2964.0 \\
    7     & 10953.9 & 3510.3 & 1319.2 & 10208.5 & 3345.7 & 1686.6 \\
    8     & 8638.8 & 3384.6 & 1950.8 & 7976.7 & 3174.4 & 1885.1 \\
    9     & 9915.3 & 2855.0 & 2100.9 & 8976.3 & 2798.3 & 2150.0 \\
    10    & 8569.3 & 3457.8 & 3259.9 & 7941.2 & 3220.6 & 3034.3 \\
    11    & 10404.3 & 3969.5 & 1935.1 & 9714.4 & 3496.9 & 2086.9 \\
    12    & 8972.4 & 3386.2 & 3383.6 & 8444.0 & 3139.5 & 3139.6 \\
    13    & 9830.7 & 1931.5 & 1931.5 & 9414.0 & 2109.4 & 2109.4 \\
    14    & 9139.1 & 3631.1 & 2471.2 & 8621.3 & 3478.1 & 2462.4 \\
    15    & 8849.3 & 3558.6 & 3442.7 & 8286.1 & 3401.1 & 3212.9 \\
    16    & 9142.5 & 3740.9 & 2490.7 & 8612.8 & 3634.1 & 2511.9 \\
    17    & 9714.6 & 1984.8 & 1984.8 & 9272.6 & 2236.4 & 2236.4 \\
    18    & 11142.9 & 3570.0 & 764.1 & 10502.8 & 3469.4 & 1499.3 \\
    19    & 10258.3 & 4006.8 & 2332.0 & 9607.2 & 3479.8 & 2567.3 \\
    20    & 9928.8 & 3506.9 & 2797.9 & 9229.8 & 3396.2 & 2748.4 \\
    \end{tabular}%
    \end{ruledtabular}
  \label{tab:1}%
\end{table}%

In order to legitimize the application of the SAA(e$\upmu$/n) framework, it must be demonstrated that muon's vibrations are sufficiently decoupled from the rest of the nuclear vibrations. To test this hypothesis, we selected a group of diverse organic molecules, depicted in Fig. \ref{fig:1}, which forms the backbone structure of many more complex organic molecules containing different potential muon and Mu binding sites. As can be seen in the figure, \textbf{1}-\textbf{11} are radicals containing a single unpaired electron, while \textbf{12}-\textbf{20} are closed-shell species. The harmonic vibrational frequencies computed at B3LYP/pc-2 level,\cite{becke_densityfunctional_1993,lee_development_1988,jensen_polarization_2001,jensen_polarization_2002} are shown in Fig. \ref{fig:2} and gathered in Tables \ref{tab:1} and S1 in the supplementary material. For each system, there are three harmonic normal modes with reduced masses in the range of 0.10-0.16 amu, see Table S2 in the supplementary material. This range of reduced masses is quite similar to the muon's mass (0.1134 amu) and indeed the animation of these modes reveals the sole dominance of muon's vibrations; {therefore}, one may conclude that muonic normal modes are decoupled from the rest of normal modes. Similar vibrationally decoupled states have been observed for the muon-phonon coupling in solid-state crystalline systems where the muon and some of its adjacent atoms may be conceived as a "molecule-in-a-crystal" defect.\cite{moller_quantum_2013,moller_playing_2013} 

\begin{figure}[t]
\includegraphics[width=\columnwidth]{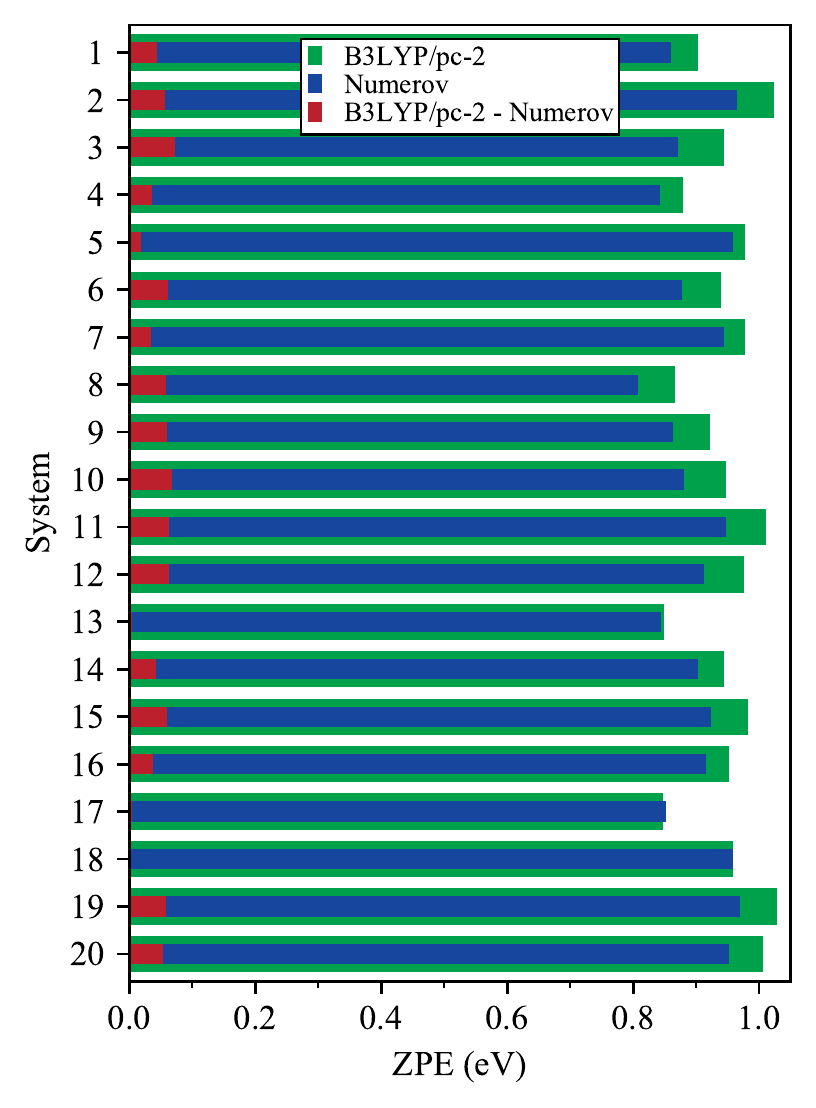}%
\caption{The ZPEs of muon vibrations computed for the benchmark systems at B3LYP/pc-2 and the Numerov levels as well as the absolute difference between {the ZPEs obtained by} the two methods (in the case of B3LYP/pc-2 level the ZPEs are the sum of contributions from the three muonic normal modes, see text for details).}
\label{fig:3}
\end{figure}

At the next stage of analysis, based on the DAA framework, the muonic vibrational energies and eigenfunctions were computed by solving the single-particle Schr\"{o}dinger equation in a three-dimensional, 3D, cubic grid, centered at the position of the surrogate hydrogen atom at the equilibrium geometry. The energy at each grid point is computed at B3LYP/pc-2 level to ensure consistency throughout the paper; the details of the used numerical grid are presented in Table S3 in the supplementary material. The {generalized matrix} Numerov method was used to solve the single-particle Schr\"{o}dinger equation of muon while the spatial spacing of 3D grids was set {such that} the error in the \black{ground state energy} falls below $O(10^{-5})$ a.u. {using the 3D 7-point stencil}. {The original Numerov method was extended to multi-dimensional problems through matrix reformulation providing the ability to systematically improve the accuracy of the solutions and a highly sparse representation of the molecular Hamiltonian.\cite{dongjiao_generalized_2014,kuenzer_pushing_2016} The high efficiency and accuracy of the reformulated Numerov method have recently been shown in the study of vibrational modes in molecular and solid-state systems.\cite{kuenzer_probing_2018,kuenzer_four-dimensional_2019,schuler_solvation_2020} In addition, the JADAMILU eigenvalue problem library was adopted for the diagonalization procedure which takes advantage of the high sparsity level of the Hamiltonian, formed by the Numerov approach, to reduce the \black{computational} and data-storage requirements.\cite{bollhofer_jadamilu_2007}} Three 1D s-type Gaussian functions were then fitted to the derived muonic vibrational ground state eigenfunction to estimate muon's vibrational frequencies for each system. The directions were chosen in a manner that the 1D wavefunction slices were parallel or perpendicular to the vector that links the muon binding site to the maximum of muon density. These three frequencies as well as the harmonic frequencies computed at B3LYP/pc-2 level are presented in Table \ref{tab:1} while corresponding muonic ZPEs are depicted in Fig. \ref{fig:3}. These two sets of frequencies and corresponding ZPEs conform well although those computed at B3LYP/pc-2 level are in general slightly larger, which probably stem from the fact that anharmonic vibrational effects are absent from this level of calculations. This observation demonstrates \black{that} the vibrational frequencies computed within the SAA(e/$\upmu$n) and DAA frameworks are not disparate, further justifying the decoupling of muon's vibrations from the rest of nuclear vibrations in the studied molecular benchmark set. Therefore, the muonic vibrational energies and wavefunctions derived by the Numerov method is used as the reference data for subsequent parameterization of the electron-muon correlation functional as also was done in the case of parameterization of the electron-proton correlation functional.\cite{yang_development_2017,tao_multicomponent_2019}

\begin{figure}[t]
\includegraphics[width=\columnwidth]{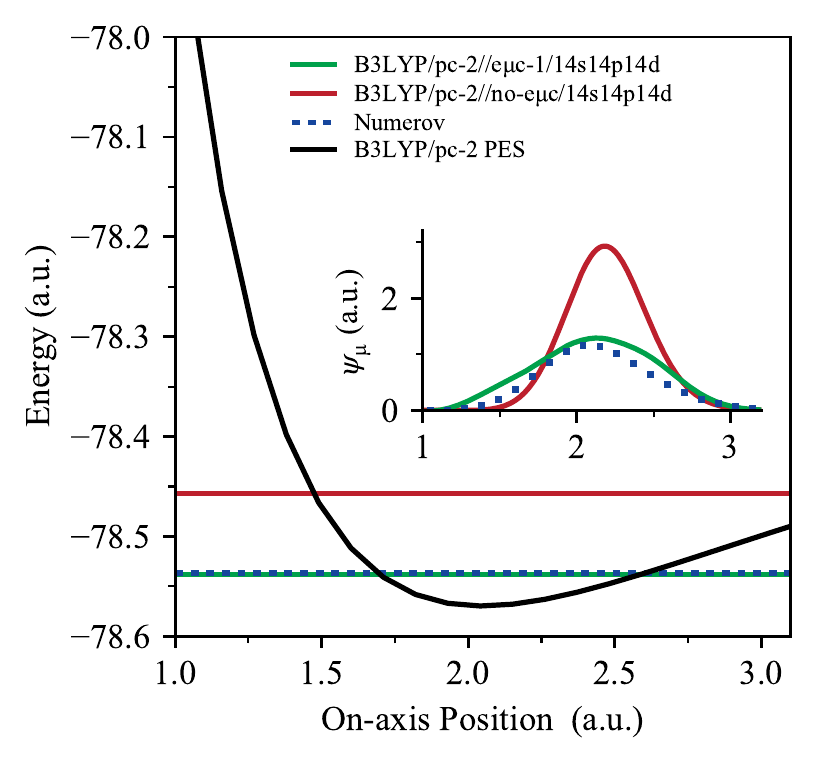}%
\caption{The B3LYP/pc-2 computed potential energy surface, PES, along the on-axis vector, total energies and muonic KS spatial orbital of \textbf{14} obtained at the TC-DFT and Numerov levels. The horizontal lines indicate the TC-DFT derived total energies and the vibrational ground state energy calculated by the Numerov method (see text for details). The inset panel shows the on-axis representation of the muonic KS orbital computed at the TC-DFT levels compared to the muonic vibrational ground state wavefunction \black{obtained} at the Numerov level. The on-axis vector goes through the center of the muonic basis set and the binding carbon atom nucleus, which is placed at the center of the coordinate system.}
\label{fig:4}
\end{figure}

\begin{figure}[ht]
\includegraphics[width=\columnwidth]{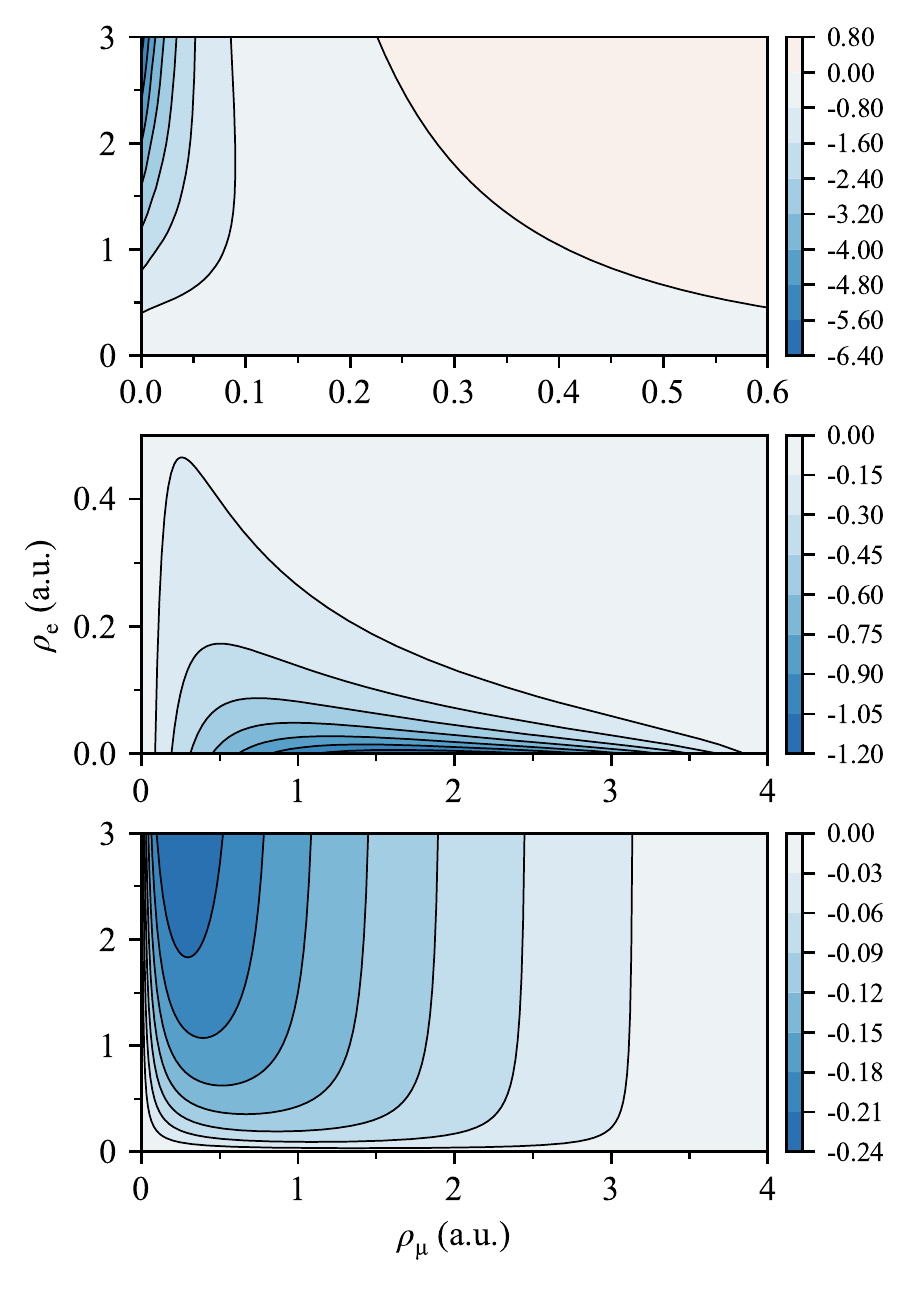}%
\caption{The contour maps of the electron-muon correlation part of the KS potentials for muon {$\nu_{\upmu}$} (top), electrons {$\nu_{\mathrm{e}}$} (middle), and the {minus of} \emc-1 kernel {$W$} (bottom) as a function of electron and muon densities (see Eqs. \ref{eq:closed_shell_EMC} and \ref{eq:ks_potentials} in Appendix \ref{sec:appendix-A}).}
\label{fig:5}
\end{figure}

\subsection{Muonic two-component density functional theory}

In this section, we will briefly review the main ingredients of the TC-DFT formalism, which is our method of choice to be applied to muonic systems. For full details the interested readers may consult the original literature.\cite{shigeta_density_1998,udagawa_hd_2006,imamura_colle-salvetti-type_2008,pak_density_2007,chakraborty_development_2008,imamura_extension_2009,chakraborty_erratum_2011,sirjoosingh_derivation_2011,chakraborty_properties_2009,sirjoosingh_multicomponent_2012,udagawa_electron-nucleus_2014,culpitt_multicomponent_2016,yang_development_2017,brorsen_alternative_2018,yang_multicomponent_2018,tao_multicomponent_2019,mejia-rodriguez_multicomponent_2019,xu_full-quantum_2020,hashimoto_analysis_2016,udagawa_nuclear_2016,udagawa_unusual_2017,brorsen_multicomponent_2017} The Hamiltonian operator of a two-component system containing a muon, $N_\mathrm{e}$ electrons and $N_\mathrm{c}$ clamped nuclei is expressed in atomic units, {a.u.}, as follows:  
\begin{align}\label{eq:hamiltonian}
\hat{H} = &-\frac{1}{2}\sum_{i=1}^{N_\mathrm{e}}\nabla_{\mathrm{e},i}^2-\frac{1}{2m_\upmu}\nabla_\upmu^2-\sum_{i=1}^{N_\mathrm{e}}\sum_{A=1}^{N_\mathrm{c}}\frac{Z_A}{|\br_{\mathrm{e},i}-\bR_A|} \nonumber\\ 
&+\sum_{A=1}^{N_\mathrm{c}}\frac{Z_A}{|\br_{\upmu}-\bR_A|}+\sum_{i=1}^{N_\mathrm{e}}\sum_{j>i}^{N_\mathrm{e}}\frac{1}{|\br_{\mathrm{e},i}-\br_{\mathrm{e},j}|}\nonumber\\
&-\sum_{i=1}^{N_\mathrm{e}}\frac{1}{|\br_{\mathrm{e},i}-\br_{\upmu}|}+\sum_{A=1}^{N_\mathrm{c}}\sum_{B>A}^{N_\mathrm{c}}\frac{Z_A Z_B}{|\bR_{A}-\bR_{B}|}.
\end{align}
In this Hamiltonian, the first two terms are the kinetic energy operators for electrons and muon, respectively, and the remaining terms in their order of appearance represent the electron-nucleus, muon-nucleus, electron-electron, electron-muon and nucleus-nucleus interaction potential energies. The electron, muon and clamped nucleus position vectors are denoted by $\br_{\mathrm{e}}$, $\br_{\upmu}$ and $\bR$, respectively, while $m_\upmu$ is the mass of muon {(206.76828 a.u.)} and $Z_A$ is the charge of the {$A$th} clamped nucleus. {Similar to the Kohn-Sham, KS, approach in the e-DFT,\cite{koch_chemists_2001,kohanoff_electronic_2006,engel_density_2011} one may assume the existence of a non-interacting two-component system with the same one-particle densities for electrons and muon in the fully-interacting two-component system. Accordingly, the underlying wavefunction of such non-interacting system is given by a simple product of the Slater determinant for electrons and the muonic spin-orbital. The total energy functional of the above two-component system in terms of the one-particle electron, $\erho$, and muon, $\mrho$, densities, based on the Hamiltonian in Eq. \ref{eq:hamiltonian} {for a closed-shell system of electrons}, is expressed by
\begin{align}\label{eq:total_energy}
E[\erho,\mrho] = &T_{\mathrm{e}}^{s}[\erho]+T_\upmu^{s}[\mrho] \nonumber\\
&+V_{\mathrm{e}}[\erho]+V_\upmu[\mrho] \nonumber\\
&+J_{\mathrm{ee}}[\erho]+J_{\mathrm{e\upmu}}[\erho,\mrho] \nonumber\\
&+E_{\mathrm{exc}}[\erho,\black{\mrho}]+E_{\mathrm{e\upmu c}}[\erho,\mrho],
\end{align}
where the non-interacting kinetic energy functionals for electrons and muon in terms of the KS spatial orbital of electrons, $\psi_{\mathrm{e}}$, and the muonic spatial orbital, $\psi_\upmu$, are given by
\begin{align}
T_{\mathrm{e}}^{s}[\erho] &= -\sum_{i=1}^{N_\mathrm{e}/2} \int \mathrm{d} \br_{\mathrm{e}} \psi_{\mathrm{e},i}^{*} (\br_{\mathrm{e}}) \nabla_{\mathrm{e}}^2 \psi_{\mathrm{e},i} (\br_{\mathrm{e}}), \nonumber\\
T_\upmu^{s}[\mrho] &= -\frac{1}{2m_\upmu} \int \mathrm{d} \br_\upmu \psi_\upmu^{*} (\br_{\upmu}) \nabla_{\upmu}^2 \psi_\upmu (\br_{\upmu}),
\end{align}
respectively. Furthermore, The electronic and muonic external potential energies are given by
\begin{align}
V_{\mathrm{e}}[\erho] &= -\sum_{A=1}^{N_\mathrm{c}}Z_A \int \mathrm{d} \br_{\mathrm{e}} \frac{\erho(\br_{\mathrm{e}})}{|\br_{\mathrm{e}}-\bR_A|}, \nonumber\\
V_\upmu[\mrho] &= \sum_{A=1}^{N_\mathrm{c}}Z_A \int \mathrm{d} \br_\upmu \frac{\mrho(\br_\upmu)}{|\br_{\upmu}-\bR_A|},
\end{align}
respectively, while the classical Coulomb electron-electron and electron-muon potential energy functionals are as follows:
\begin{align}
J_{\mathrm{ee}}[\erho] &=  \frac{1}{2} \int \mathrm{d} \br_{\mathrm{e}} \int \mathrm{d} \br'_{\mathrm{e}} \frac{\erho(\br_{\mathrm{e}}) \erho(\br'_{\mathrm{e}})}{|\br_{\mathrm{e}}-\br'_{\mathrm{e}}|},  \nonumber\\
J_{\mathrm{e\upmu}}[\erho,\mrho] &= - \int \mathrm{d} \br_{\mathrm{e}} \int \mathrm{d} \br_\upmu \frac{\erho(\br_{\mathrm{e}}) \mrho(\br_\upmu)}{|\br_{\mathrm{e}}-\br_{\upmu}|},
\end{align}
respectively. The last two terms in Eq. \ref{eq:total_energy} are the electron exchange-correlation energy functional and the electron-muon correlation energy functional, which are the main unknowns, and can be defined as
\begin{align}\label{eq:functionals}
E_{\mathrm{exc}}[\erho,\black{\mrho}] = & (T_{\mathrm{e}}[\erho,\black{\mrho}]-T_{\mathrm{e}}^{s}[\erho]) + (V_{\mathrm{ee}}[\erho,\black{\mrho}]-J_{\mathrm{ee}}[\erho]), \nonumber \\
E_{\mathrm{e\upmu c}}[\erho,\mrho]  = & (T_{\mathrm{\upmu}}[\black{\erho},\mrho]-T_\upmu^{s}[\mrho]) \nonumber\\
&+ (V_{\mathrm{e\upmu}}[\erho,\mrho]-J_{\mathrm{e\upmu}}[\erho,\mrho]),
\end{align}
respectively. The electronic kinetic energy functional and the electron-electron interaction energy functional are given by $T_{\mathrm{e}}[\erho,\black{\mrho}]$ and $V_{\mathrm{ee}}[\erho,\black{\mrho}]$, respectively. Moreover, the electron-muon interaction energy functional is defined as $V_{\mathrm{e\upmu}}[\erho,\mrho]$, while $T_{\mathrm{\upmu}}[\black{\erho},\mrho]$ is the muonic kinetic energy functional.} 

The minimization of the total energy functional Eq. \ref{eq:total_energy} with respect to the electronic and muonic spin-orbitals, subject to the orthonormalization constraint of spin-orbitals and assuming a closed electronic shell, leads to a set of coupled eigenvalue equations, {that are} the two-component KS equations for electrons and the muon:
\begin{align}\label{eq:ks_equations}
\hat{h}_{\mathrm{e}} \psi_{\mathrm{e},i} (\br_{\mathrm{e}}) &= \varepsilon_{\mathrm{e},i} \psi_{\mathrm{e},i} (\br_{\mathrm{e}}),  i=1, \dots, N_\mathrm{e}/2, \nonumber \\
\hat{h}_\upmu \psi_\upmu (\br_{\upmu}) &= \varepsilon_{\upmu} \psi_\upmu (\br_{\upmu}).
\end{align}
In the above equations, $\varepsilon_{\mathrm{e},i}$ and $\varepsilon_{\upmu}$ are the KS orbital energies for the $i$th KS spatial orbital of electrons, $\psi_{\mathrm{e},i}$, and the muonic spatial orbital, $\psi_\upmu$, respectively, while {the effective KS one-particle Hamiltonian} operators of electrons and muon are as follows:
\begin{align}
\hat{h}_{\mathrm{e}}(\br_{\mathrm{e}}) =& -\frac{1}{2}\nabla_{\mathrm{e}}^2-\sum_{A=1}^{N_\mathrm{c}}\frac{Z_A}{|\br_{\mathrm{e}}-\bR_A|} \nonumber\\
& -\int \mathrm{d} \br_\upmu \frac{\mrho(\br_\upmu)}{|\br_{\mathrm{e}}-\br_{\upmu}|}+\int \mathrm{d} \br'_{\mathrm{e}} \frac{\erho(\br'_{\mathrm{e}})}{|\br_{\mathrm{e}}-\br'_{\mathrm{e}}|}  \nonumber \\
& +\frac{\delta E_{\mathrm{exc}}[\erho,\black{\mrho}]}{\delta \erho}+\frac{\delta E_{\mathrm{e\upmu c}}[\erho,\mrho]}{\delta \erho}, \nonumber \\
\hat{h}_\upmu(\br_{\upmu}) =& -\frac{1}{2m_\upmu}\nabla_{\upmu}^2+\sum_{A=1}^{N_\mathrm{c}}\frac{Z_A}{|\br_{\upmu}-\bR_A|} \nonumber\\
& -\int \mathrm{d} \br_{\mathrm{e}} \frac{\erho(\br_{\mathrm{e}})}{|\br_{\upmu}-\br_{\mathrm{e}}|}
+\frac{\delta E_{\mathrm{e\upmu c}}[\erho,\mrho]}{\delta \mrho}.
\end{align}
{The above} operators contain the kinetic energy, the external potentials and the two-particle classical Coulomb interactions, which are explicitly known. The remaining interaction potentials, {that are} the electron exchange-correlation potential and the electron-muon correlation potential, {as mentioned previously,} are generally unknown and proper approximations should be made in practice to allow for accurate computational results. It has been argued that the conventional electronic exchange-correlation density functional approximations, used within the context of the e-DFT, can be used without reparameterization for the two-component protonic systems.\cite{sirjoosingh_multicomponent_2012,brorsen_alternative_2018} Although there are some hesitations,\cite{udagawa_electron-nucleus_2014} this seems to be an appropriate approximation in first developmental attempts in the case of muonic systems. Therefore herein, we follow the same line of reasoning and mainly focus on the development of approximations for the electron-muon correlation \black{functional}{, as given in Eq. \ref{eq:functionals}}. 

The coupled set of KS equations in Eq. \ref{eq:ks_equations} may be solved iteratively, through the self-consistent field, SCF, procedure, starting from an initial guess of the electronic and muonic KS spatial orbitals, until reaching a convergent state. Based on the two-component KS wavefunction, the electron and muon densities are defined by their corresponding KS orbitals as follows:
\begin{align}
\erho &=2 \sum_{i=1}^{N_\mathrm{e}/2} |\psi_{\mathrm{e},i} (\br_{\mathrm{e}})|^2, \nonumber \\
\mrho &= |\psi_\upmu (\br_{\upmu})|^2.
\end{align}
In the case of the electronically spin-polarized open-shell systems, a similar variational procedure is used and the energy minimization leads to two sets of coupled KS equations for alpha and beta electrons, which upon the SCF convergence, provides their respective one-electron densities:
\begin{align}
\arho &= \sum_{i=1}^{N_\upalpha} |\psi_{\upalpha,i} (\br_{\upalpha})|^2, \nonumber \\
\brho &= \sum_{i=1}^{N_\upbeta} |\psi_{\upbeta,i} (\br_{\upbeta})|^2,
\end{align}
where $N_\upalpha$ and $N_\upbeta$ are the numbers of alpha and beta electrons, respectively.

\subsection{Development of the electron-muon correlation functional}

The electron-muon correlation functional is the key ingredient of the muonic TC-DFT that to the best of our knowledge has not been addressed so far in {the} literature. In this section, we will present our proposal for a reliable, but simple local approximate functional. We start this section with a brief review of the recently proposed electron-proton correlation functionals as a relevant development. To date, several groups have made attempts to develop electron-proton correlation functionals for protonic systems.\cite{shigeta_density_1998,udagawa_hd_2006,pak_density_2007,imamura_colle-salvetti-type_2008,chakraborty_development_2008,chakraborty_properties_2009,imamura_extension_2009,chakraborty_erratum_2011,sirjoosingh_derivation_2011,sirjoosingh_multicomponent_2012,udagawa_electron-nucleus_2014,culpitt_multicomponent_2016,hashimoto_analysis_2016,udagawa_nuclear_2016,udagawa_unusual_2017,brorsen_multicomponent_2017,yang_development_2017,brorsen_alternative_2018,yang_multicomponent_2018,tao_multicomponent_2019,mejia-rodriguez_multicomponent_2019,xu_full-quantum_2020} One way of deducing the correlation functional is to use the known many-body results \black{for} two-component systems, for example various two-component homogeneous electron-positively charged particle gases, as has been done in the case of protonic and also positronic systems.\cite{boronski_electron-positron_1986,ito_formulation_2004} Another line of research is based on extending the original Colle-Salvetti approach for approximating the electronic exchange-correlation energy density through the electron density and local electronic kinetic energy density.\cite{colle_approximate_1975,lee_development_1988} Accordingly, the electron-proton correlation energy density may be approximated locally based on the electron and proton densities.\cite{udagawa_hd_2006, imamura_colle-salvetti-type_2008,imamura_extension_2009,udagawa_electron-nucleus_2014,yang_development_2017,brorsen_alternative_2018} One of the computationally successful members of this family is the epc17 functional:
\begin{equation}
E_{\mathrm{epc17}}[\erho,\prho] = -\int \mathrm{d} \br \frac{\erho \prho} {a - b \sqrt{\erho\prho} + c \erho \prho}.
\end{equation}
\black{In this functional, $\prho$ is the proton density and $\int \mathrm{d} \br \equiv \int \mathrm{d} \br_{\mathrm{e}} \int \mathrm{d} \br_{\mathrm{p}} \delta(\br_{\mathrm{e}}-\br_{\mathrm{p}})$ where $\br_{\mathrm{p}}$ is the proton position vector. Furthermore, $a$, $b$ and $c$ are parameters which are determined by a {fitting} procedure.\cite{brorsen_multicomponent_2017}} More recently a gradient-corrected extension of the epc17 functional, termed as epc19, also has been introduced.\cite{tao_multicomponent_2019} A more rigorous way of developing the correlation functional is achieved through using the explicitly-correlated electron-proton wavefunctions.\cite{chakraborty_development_2008,chakraborty_erratum_2011,sirjoosingh_derivation_2011} However, the general computation implementation of these functionals seems to be quite demanding prohibiting their applications to large molecular systems. Probably in the long term, it is more desirable to design the electron-positively charged particle correlation functional {by} employing the properties of the exact functional that are known analytically.\cite{chakraborty_properties_2009} On the other hand, the success of epc17 encourages adopting a more semi-empirical approach that seems to be acceptable at the moment, as was the case in the initial phase of the development of the e-DFT.\cite{koch_chemists_2001,kohanoff_electronic_2006,engel_density_2011} Based on this background, our goal is to propose an electron-muon correlation functional that satisfies the following requirements:
\begin{enumerate}
    \item To be capable of reproducing the muon density and muonic kinetic energy of the muonic system in an accurate quantitative manner. 
    \item To be capable of reproducing the potential energy surface of muonic systems; yielding corresponding optimized geometries as well as the vibrational spectrum of clamped nuclei without referring to any prior calculations within the DAA and SAA(e/$\upmu$n) frameworks.
\end{enumerate}
We believe if these two requirements are {met} the proposed electron-muon correlation functional yields a self-sufficient TC-DFT(ee+e$\upmu$) method which can be implemented computationally and applied to large muonic molecules.

\begin{table*}[t]
	\centering
	\caption{The total energies ($E_h$), muonic kinetic energies ($E_h$), muonic bond length expectation values (\r{A}) and RMSDs of muon densities (a.u.) calculated via the TC-DFT and Numerov methods.}
	\begin{ruledtabular}
	\begin{tabular}{l d{3.4}d{1.4}d{1.3} d{3.4}d{1.4}d{1.3}d{1.3} d{3.4}d{1.4}d{1.3}d{1.3}}%
		& \multicolumn{3}{c}{Numerov} & \multicolumn{4}{c}{B3LYP/pc-2//\emc-1/14s14p14d} & \multicolumn{4}{c}{B3LYP/pc-2//no-\emc/14s14p14d} \\
		\cmidrule{2-4} \cmidrule{5-8} \cmidrule{9-12}
		 System & \multicolumn{1}{c}{$E$\footnotemark[1]} & \multicolumn{1}{c}{$K_{\upmu}$\footnotemark[2]} & \multicolumn{1}{c}{$\langle r_{\upmu} \rangle$}& 
		\multicolumn{1}{c}{$E$\footnotemark[3]} & \multicolumn{1}{c}{$K_{\upmu}$} & \multicolumn{1}{c}{$\langle r_{\upmu} \rangle$}& \multicolumn{1}{c}{RMSD\footnotemark[4]}&
		\multicolumn{1}{c}{$E$\footnotemark[3]} & \multicolumn{1}{c}{$K_{\upmu}$} & \multicolumn{1}{c}{$\langle r_{\upmu} \rangle$}& \multicolumn{1}{c}{RMSD\footnotemark[4]}\\ \hline
    1     & -77.8548 & 0.0157 & 1.092 & -77.8551 & 0.0192 & 1.100 & 0.031 & -77.7751 & 0.0451 & 1.157 & 0.154 \\
    2     & -111.1890 & 0.0176 & 0.998 & -111.1993 & 0.0186 & 1.004 & 0.035 & -111.1138 & 0.0445 & 1.071 & 0.152 \\
    3     & -79.1028 & 0.0163 & 1.129 & -79.1037 & 0.0192 & 1.118 & 0.021 & -79.0241 & 0.0452 & 1.176 & 0.146 \\
    4     & -115.0143 & 0.0154 & 1.133 & -115.0143 & 0.0192 & 1.127 & 0.030 & -114.9362 & 0.0438 & 1.192 & 0.147 \\
    5     & -115.0224 & 0.0170 & 0.918 & -115.0392 & 0.0180 & 0.936 & 0.039 & -114.9493 & 0.0428 & 1.023 & 0.155 \\
    6     & -170.3688 & 0.0162 & 1.136 & -170.3695 & 0.0193 & 1.130 & 0.022 & -170.2913 & 0.0440 & 1.198 & 0.145 \\
    7     & -170.3808 & 0.0170 & 0.933 & -170.3977 & 0.0180 & 0.941 & 0.035 & -170.3086 & 0.0423 & 1.029 & 0.151 \\
    8     & -93.9406 & 0.0149 & 1.107 & -93.9395 & 0.0192 & 1.115 & 0.036 & -93.8608 & 0.0443 & 1.175 & 0.155 \\
    9     & -93.9265 & 0.0159 & 1.019 & -93.9341 & 0.0186 & 1.015 & 0.031 & -93.8505 & 0.0432 & 1.088 & 0.149 \\
    10    & -95.1440 & 0.0164 & 1.127 & -95.1451 & 0.0192 & 1.120 & 0.023 & -95.0658 & 0.0449 & 1.179 & 0.146 \\
    11    & -95.1543 & 0.0173 & 0.992 & -95.1645 & 0.0186 & 1.002 & 0.035 & -95.0786 & 0.0445 & 1.071 & 0.154 \\
    12    & -40.4707 & 0.0169 & 1.109 & -40.4729 & 0.0192 & 1.104 & 0.019 & -40.3923 & 0.0458 & 1.157 & 0.147 \\
    13    & -77.2870 & 0.0150 & 1.039 & -77.2890 & 0.0189 & 1.069 & 0.040 & -77.2080 & 0.0441 & 1.128 & 0.159 \\
    14    & -78.5365 & 0.0165 & 1.095 & -78.5377 & 0.0192 & 1.098 & 0.025 & -78.4571 & 0.0459 & 1.151 & 0.151 \\
    15    & -79.7692 & 0.0171 & 1.116 & -79.7711 & 0.0193 & 1.109 & 0.018 & -79.6905 & 0.0461 & 1.161 & 0.146 \\
    16    & -232.1577 & 0.0168 & 1.096 & -232.1593 & 0.0192 & 1.098 & 0.024 & -232.0785 & 0.0460 & 1.151 & 0.150 \\
    17    & -93.3853 & 0.0147 & 1.046 & -93.3882 & 0.0189 & 1.072 & 0.039 & -93.3077 & 0.0432 & 1.135 & 0.156 \\
    18    & -115.6816 & 0.0168 & 0.900 & -115.6980 & 0.0180 & 0.932 & 0.043 & -115.6073 & 0.0433 & 1.020 & 0.159 \\
    19    & -95.8060 & 0.0178 & 1.011 & -95.8164 & 0.0186 & 1.007 & 0.025 & -95.7303 & 0.0451 & 1.074 & 0.148 \\
    20    & -94.5820 & 0.0175 & 1.032 & -94.5908 & 0.0187 & 1.023 & 0.021 & -94.5063 & 0.0448 & 1.087 & 0.144 \\
        ME\footnotemark[5]    & & & &  -0.0055 & 0.0024 & 0.005 & 0.030 & 0.0771 & 0.0280 & 0.070 & 0.151 \\
        RMSE\footnotemark[6]  & & & &  0.0081 & 0.0027 & 0.014 & & 0.0771 & 0.0280 & 0.073 & \\
        
	\end{tabular}%
    \end{ruledtabular}
    \footnotetext[1]{The total energy is equal to the sum of the equilibrium energy obtained at B3LYP/pc-2 level and the muonic ZPE, that is the muonic vibrational ground state energy computed by the Numerov method.}
    \footnotetext[2]{The kinetic energy is computed as half of the muonic ZPE, assuming the validity of the virial theorem for the harmonic oscillator model.}
    \footnotetext[3]{The TC-DFT energy contains the electronic and muonic energy contributions as quantum particles (see Eq. \ref{eq:total_energy}).}
    \footnotetext[4]{The RMSD of muon densities is obtained as the square root of the average of squared difference between the TC-DFT and Numerov derived densities, calculated over the entire 3D grid used for the Numerov calculations (see Table S3 in the supplementary material).}
    \footnotetext[5]{The mean error, ME, is defined as the average of the deviations of the TC-DFT predictions from the Numerov reference data.}
    \footnotetext[6]{The root mean square error, RMSE, is defined as the square root of the average of squared deviations of the TC-DFT predictions from the Numerov reference data.}
	\label{tab:2}%
\end{table*}%

In the present study, by employing the general idea of local Wigner-type functionals used in the e-DFT,\cite{wigner_interaction_1934,wilson_nonlocal_1990,zhao_local_1992,wilson_development_1994,stewart_beckewigner_1995} and through a proper generalization for the two-component systems, we introduce a local electron-muon correlation functional which is capable of overcoming the overlocalization problem previously discussed. The general forms of Pad\'{e} approximant of order $m$ and $n$ for the electronic functionals,\cite{lopez-boada_pade_1997} and its tentative extension to the two-component muonic systems are as follows:  
\begin{align}
E_{\mathrm{exc}}^{m,n}[\erho] &=  \int \mathrm{d} \br_{\mathrm{e}} \frac{\sum_{i=0}^{m}{a_i \erho^{p_i}}}{\sum_{i=0}^{n}{b_i \erho^{l_i}}}, \nonumber\\
E_{\mathrm{e\upmu c}}^{m,n}[\erho,\mrho] &= - \int \mathrm{d} \br \frac{\sum_{i=0}^{m}{a_i \erho^{p_i} \mrho^{q_i}}}{\sum_{i=0}^{n}{b_i \erho^{l_i} \mrho^{k_i}}},
\end{align}
where all linear coefficients and exponents of densities are parameters to be determined from theoretical models or regression procedures while $\int \mathrm{d} \br \equiv \int \mathrm{d} \br_{\mathrm{e}} \int \mathrm{d} \br_{\upmu} \delta(\br_{\mathrm{e}}-\br_{\upmu})$; in the present study, the latter approach is used to deduce the proper set of parameters. Herein, we used the Numerov method results discussed in the previous section, as the reference data, to acquire the {\emc} parameters by fitting to a particular system and then evaluating the derived functional performance in the set of benchmark molecular systems presented in Fig. \ref{fig:1}.

The fitting procedure was done using \textbf{14} \black{where} the total energy, muonic kinetic energy and muonic bond length expectation value were selected as the primary target to minimize the difference to their corresponding Numerov results while the overall shape of the muonic KS spatial orbital was also considered as a secondary measure. The TC-DFT(ee+e$\upmu$) calculations were performed with B3LYP as the {electronic} exchange-correlation functional using {spherical} pc-2 {basis functions} as the electronic basis set {where the standard spherical pc-2 basis set for hydrogen was also employed for the Mu atom}. The used muonic basis set consists of 14s14p14d {Cartesian} Gaussian functions, located at a single center in 3D space, whose exponents {form} an even-tempered series {derived from} $2(\sqrt{2})^{(i-3)}$ formula, where $i$ is a non-negative integer. The selected exponent range starts from $\sqrt{2}/2$ to 64, corresponding to $i=\textrm{0-13}$, which makes the basis set flexible enough to accurately reproduce the muon density in various chemical environments, {and removes the need for further basis set exponent optimizations}. The geometry of the target system, that is the positions of the clamped nuclei as well as the center of the muonic basis set, was fully optimized during the parameterization procedure. The final optimized form of the electron-muon correlation functional, hereafter denoted as \emc-1, is as follows:
\begin{align}\label{eq:closed_shell_EMC}
E_{\mathrm{e\upmu c}}[\erho,\mrho] &= -\int \mathrm{d} \br W[\erho,\mrho],\nonumber \\
W[\erho,\mrho] &= \frac{2 \erho \mrho - \erho (\mrho)^{3/2}} {1 + 4\erho\mrho + 2\erho (\mrho)^{3/2}}.
\end{align}
The \emc-1 energy kernel, which has the physical dimension of energy per volume, does not have any singularity, as opposed to some of previously reported electron-nucleus correlation functionals.\cite{imamura_colle-salvetti-type_2008,udagawa_electron-nucleus_2014}

Let us consider the performance of \emc-1 for \textbf{14} in more detail. The total energies, and 1D slices of {3D} normalized muonic spatial orbitals at B3LYP/pc-2//\emc-1/14s14p14d and B3LYP/pc-2//no-\emc/14s14p14d levels are shown in Fig. \ref{fig:4}, and compared to the Numerov reference results. By introducing the \emc-1 functional, the TC-DFT derived total energy and the Numerov vibrational ground-state energy are almost identical while the difference of these two with B3LYP/pc-2//no-\emc/14s14p14d derived total energy is substantial. The lack of the electron-muon correlation also {leads} to the overlocalization of the muonic KS orbital, which is a well-documented effect also seen for proton previously.\cite{yang_development_2017} The overall shape of the muon distribution derived at the Numerov level is also replicated fairly well with the inclusion of the {\emc-1} functional. The considerable overlocalization at this level also leads to a large kinetic energy component which is also improved significantly by the inclusion of the \emc-1 functional. It is easily {noticeable} that the lack of the {\emc} functional would also {result in the} overestimation of the muonic average bond length. These results are highly encouraging revealing the fact that this local functional is capable of delivering reasonable results for not only the total {energy} and muonic kinetic {energy} but also {the} muonic KS orbitals at a fraction of the computational cost required by the Numerov method. 

\begin{figure*}[t]
\includegraphics[]{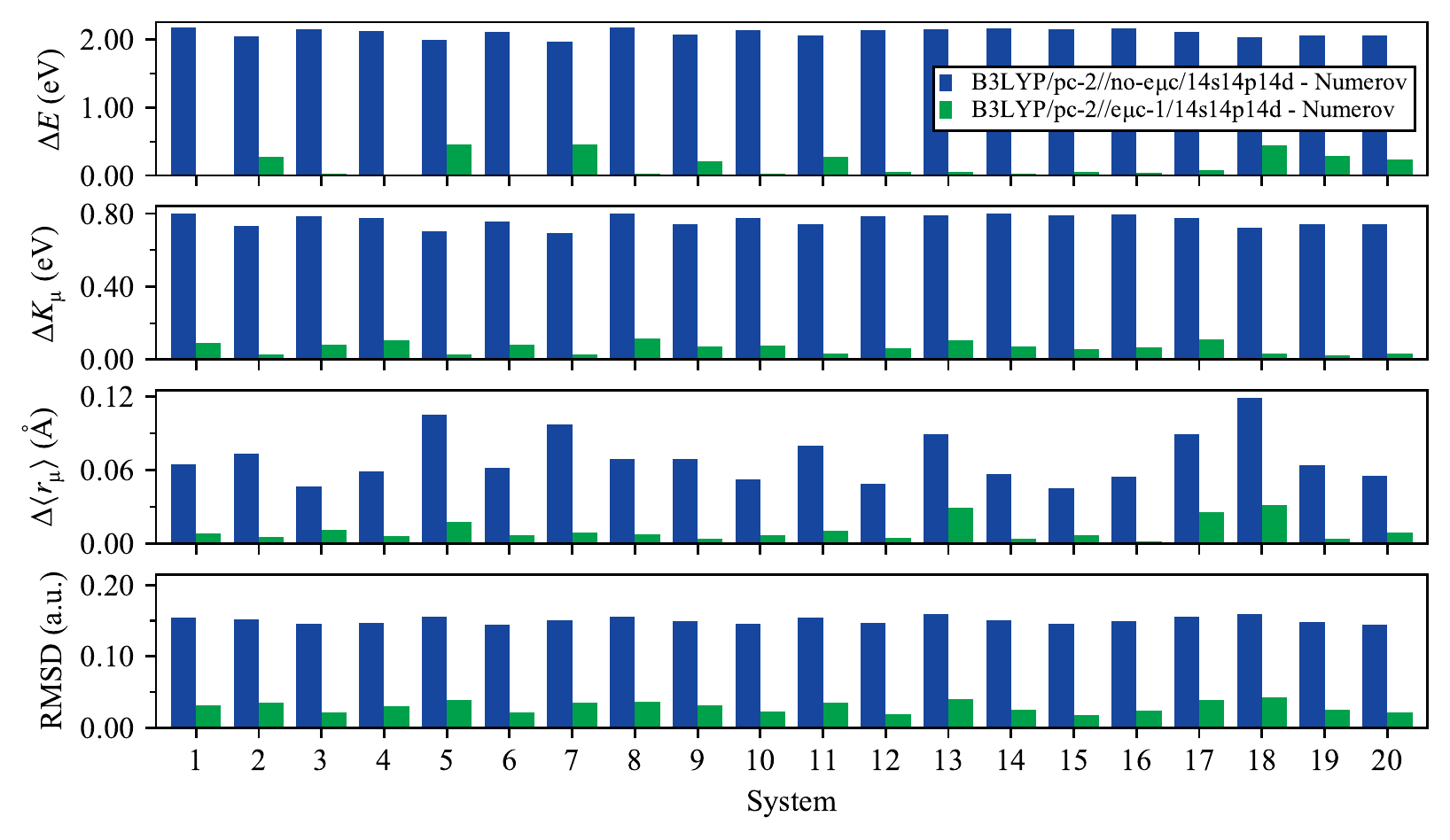}%
\caption{The absolute differences in the total energies, muonic kinetic energies, muonic bond length expectation values and RMSDs of muon densities between the TC-DFT and Numerov methods for the benchmark set of systems (see Table \ref{tab:2} for numerical values).}
\label{fig:6}
\end{figure*}

\begin{figure*}[t]
\includegraphics[width=\textwidth]{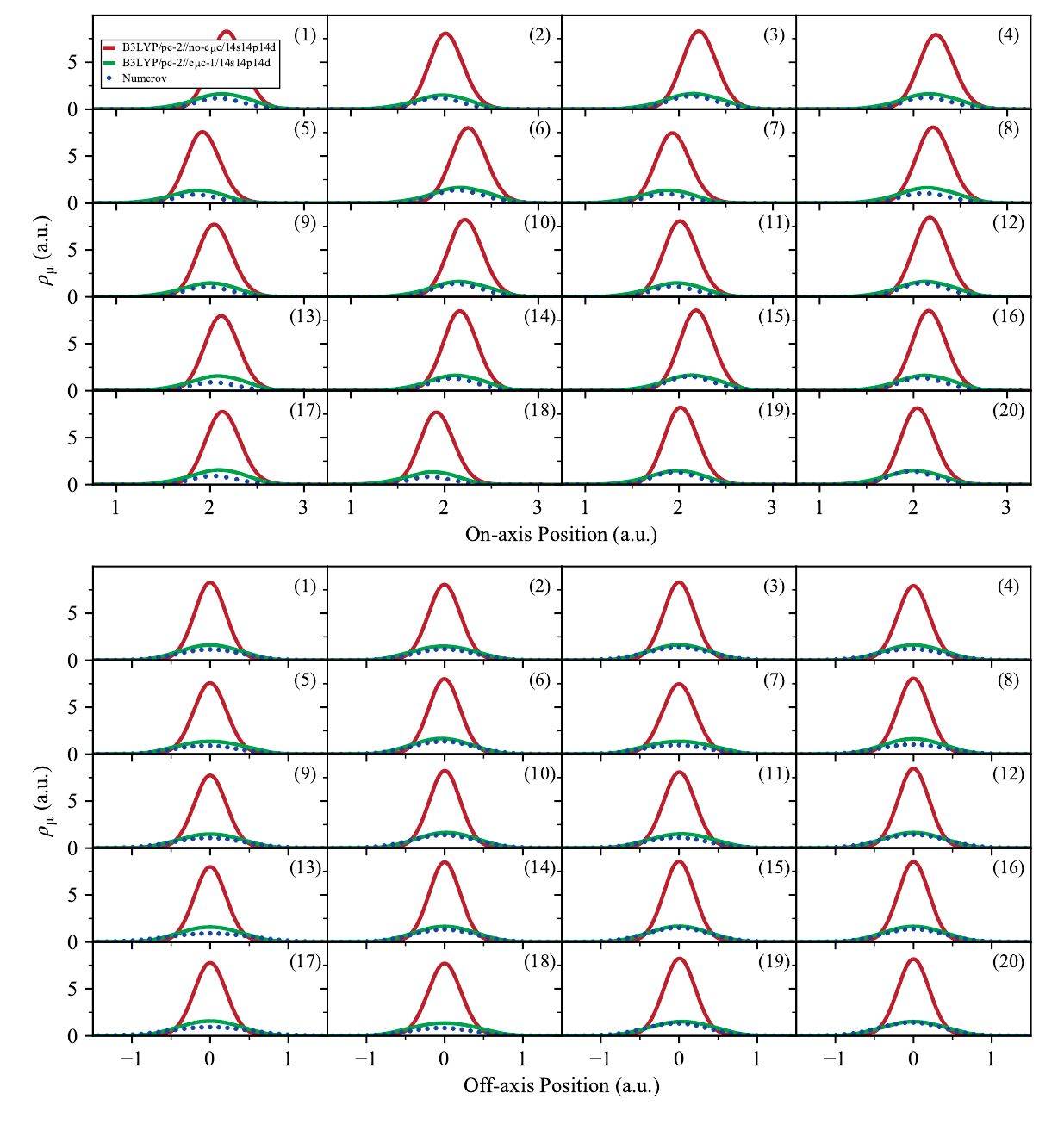}%
\caption{The on-axis (top) and the off-axis (bottom) representations of muon density computed via the TC-DFT and Numerov methods. The on-axis position indicates the direction which is parallel to the vector connecting the muon binding site to the muon density maximum obtained at the corresponding level of theory while the off-axis position refers to the slice that is perpendicular to the on-axis direction and goes through the respective muon density maximum. The muon binding site resides at the center of the coordinate system. \black{The corresponding on- and off-axis muonic vibrational frequencies derived from the underlying muonic wavefunctions at the Numerov level of theory are offered in the fifth and sixth columns of Table \ref{tab:1}.}}
\label{fig:7}
\end{figure*}

\begin{table*}[ht]
  \centering
  \caption{The sum of total energies ($E_h$) derived at the equilibrium energy and the ZPEs of clamped nuclei (computed from the harmonic normal mode frequencies gathered in Table S1), calculated at B3LYP/pc-2 and the TC-DFT levels for the benchmark set of systems.}
    \begin{ruledtabular}
    \begin{tabular}{l d{3.4}d{3.4}d{3.4} d{1.3}d{1.3}d{1.3}}
    System  & \multicolumn{1}{c}{{B3LYP/pc-2}}   & \multicolumn{1}{c}{\text{B3LYP/pc-2//\emc-1/14s14p14d}} & \multicolumn{1}{c}{{B3LYP/pc-2//no-\emc/14s14p14d}} & \multicolumn{1}{c}{$\Delta$\footnotemark[1]} & \multicolumn{1}{c}{$\Delta$\footnotemark[2]} & \multicolumn{1}{c}{$\Delta$\footnotemark[3]} \\ \hline
    1     & -77.8288 & -77.8307 & -77.7507 & -0.002 & 0.078 & 0.000 \\
    2     & -111.1610 & -111.1733 & -111.0880 & -0.012 & 0.073 & -0.010 \\
    3     & -79.0544 & -79.0572 & -78.9774 & -0.003 & 0.077 & -0.001 \\
    4     & -114.9886 & -114.9899 & -114.9117 & -0.001 & 0.077 & 0.000 \\
    5     & -114.9960 & -115.0147 & -114.9248 & -0.019 & 0.071 & -0.017 \\
    6     & -170.3248 & -170.3269 & -170.2485 & -0.002 & 0.076 & -0.001 \\
    7     & -170.3362 & -170.3551 & -170.2658 & -0.019 & 0.070 & -0.017 \\
    8     & -93.9255 & -93.9264 & -93.8478 & -0.001 & 0.078 & 0.001 \\
    9     & -93.9109 & -93.9207 & -93.8371 & -0.010 & 0.074 & -0.008 \\
    10    & -95.1062 & -95.1090 & -95.0295 & -0.003 & 0.077 & -0.001 \\
    11    & -95.1156 & -95.1280 & -95.0422 & -0.012 & 0.073 & -0.010 \\
    12    & -40.4379 & -40.4422 & -40.3615 & -0.004 & 0.076 & -0.002 \\
    13    & -77.2698 & -77.2729 & -77.1919 & -0.003 & 0.078 & -0.002 \\
    14    & -78.4966 & -78.4993 & -78.4187 & -0.003 & 0.078 & -0.001 \\
    15    & -79.7061 & -79.7096 & -79.6292 & -0.004 & 0.077 & -0.002 \\
    16    & -232.0681 & -232.0709 & -231.9902 & -0.003 & 0.078 & -0.002 \\
    17    & -93.3788 & -93.3829 & -93.3025 & -0.004 & 0.076 & -0.003 \\
    18    & -115.6413 & -115.6592 & -115.5685 & -0.018 & 0.073 & -0.016 \\
    19    & -95.7539 & -95.7665 & -95.6803 & -0.013 & 0.074 & -0.010 \\
    20    & -94.5536 & -94.5643 & -94.4799 & -0.011 & 0.074 & -0.009 \\
    \end{tabular}%
    \end{ruledtabular}
    \footnotetext[1]{The difference between the ZPE-corrected energies obtained at B3LYP/pc-2//\emc-1/14s14p14d and B3LYP/pc-2 levels.}
    \footnotetext[2]{The difference between the ZPE-corrected energies computed at B3LYP/pc-2//no-\emc/14s14p14d and B3LYP/pc-2 levels.}
    \footnotetext[3]{{The difference between total energies calculated at B3LYP/pc-2//\emc-1/14s14p14d and the Numerov levels presented in the fifth and second columns of Table \ref{tab:2}, respectively.}}
  \label{tab:3}%
\end{table*}%

The contour maps of the \emc-1 derived KS potentials for the muon and electrons as well as the kernel of the functional are all depicted in Fig. \ref{fig:5} for a relevant region of the muonic and electronic densities, using the KS potentials offered in Appendix \ref{sec:appendix-A}. The muonic KS potential clearly shows a tendency for destabilization at larger values of muon densities, overcoming the previously mentioned overlocalization effect similar to the protonic KS potential in the case of the epc17 functionals.\cite{yang_development_2017} The electronic KS potential has similar bias against the electron density localization without possessing any positive potential region in contrast to the muonic KS potential. The \emc-1 kernel generally favors lower muonic densities, while a closer look at Eq. \ref{eq:closed_shell_EMC} reveals the fact that if $\mrho>4$ a.u., which is highly unlikely under ambient experimental conditions, the kernel will change sign and becomes positive, unphysically destabilizing the muonic system. It should be noted that in modeling muonic systems under very high hydrostatic pressures this change of sign may cause the functional to fail, although treating these conditions is beyond the scope of the present study.

At this stage of development, we may formulate the functional for the case of spin-polarized electronic systems. {The spin-polarized electron-muon correlation functional is composed of two separate, but fundamentally similar, alpha and beta electron contributions. Using the scaling relationships previously offered in the development of the \black{non-interacting} kinetic energy and \black{the exact} exchange energy functionals for the spin-polarized systems in the e-DFT,\cite{oliver_spin-density_1979,perdew_generalized_1996} and Eq. \ref{eq:closed_shell_EMC}, the {\emc} kernels for alpha and beta electrons are \black{approximated} as follows:}
\begin{align}\label{eq:polarized_wigner}
E_{\mathrm{e\upmu c}}[\arho,\brho,\mrho] \black{\approx}& {(E_{\mathrm{e\upmu c}}[2\arho,\mrho] + E_{\mathrm{e\upmu c}}[2\brho,\mrho])/2} \nonumber\\
=& {-\int \mathrm{d} \br W[\arho,\brho,\mrho]}, \nonumber\\
W[\arho,\brho,\mrho] =& \frac{2\arho \mrho - \arho (\mrho)^{3/2}} {1 + 8\arho\mrho + 4\arho (\mrho)^{3/2}} \nonumber\\
& +\frac{2\brho \mrho - \brho (\mrho)^{3/2}} {1 + 8\brho\mrho + 4\brho (\mrho)^{3/2}}.
\end{align}
It is readily seen that the final spin-polarized {\emc} energy kernel, Eq. \ref{eq:polarized_wigner}, transforms to the fully-unpolarized equation, Eq. \ref{eq:closed_shell_EMC}, if $\arho = \brho$.

\subsection{Testing the performance of the developed functional}

The general robustness and transferability of the \emc-1 functional were evaluated by a comparative analysis for the entire benchmark systems depicted in Fig. \ref{fig:1}. Similar to the calculations in the previous section, B3LYP/pc-2//\emc-1/14s14p14d and B3LYP/pc-2//no-\emc/14s14p14d levels of theory were used for all members of the benchmark set. To this end, our in-house version of the nuclear-electronic orbital, NEO, GAMESS package was modified to compute the total energies and corresponding analytical first derivatives in the presence of the \emc-1 functional.\cite{schmidt_general_1993,webb_multiconfigurational_2002,pak_density_2007} Table \ref{tab:2} presents the total energies, muonic kinetic energies, muonic bond length expectation values, and root mean square deviations, RMSD, of muon densities at the fully optimized geometries obtained at their respective levels of theory compared to the results derived by the Numerov method. The optimized structures are offered in Tables S4-S23 in the supplementary material. Also, Fig. \ref{fig:6} depicts explicitly the differences of the TC-DFT predictions with the Numerov results. It is evident from Fig. \ref{fig:6} that by inclusion of the \emc-1 functional into the TC-DFT, the total {energies} and muonic kinetic energies, and the muonic bond length expectation values are all improved by an order of magnitude in both electronically closed- and open-shell systems. Furthermore, the computed muon densities which are depicted in on- and off-axis directions in Fig. \ref{fig:7}, in line with the RMSDs offered in Table \ref{tab:2}, are substantially improved upon the inclusion of the \emc-1 functional. All these indicate that the \emc-1 based TC-DFT(ee+e$\upmu$) meets the first requirement of a self-sufficient computational method.

A closer inspection of the \emc-1 included results reveals that the largest deviations from the reference Numerov results for total energies, less than 0.5 eV, are observed for the oxygen binding sites in \textbf{5}, \textbf{7}, and \textbf{18}, and nitrogen centers in \textbf{2}, \textbf{9}, \textbf{11}, \textbf{19}, and \textbf{20}. The maximum deviations for the muonic average bond lengths are for systems with oxygen binding sites and the sp-hybridized carbon centers in \textbf{13} and \textbf{17}. This is probably related to the fact that the \emc-1 functional was parameterized for an $\mathrm{sp}^{2}$ carbon attachment site in \textbf{14}. 

Table \ref{tab:3} contains the ZPE-corrected energies where the ZPEs have been computed from the harmonic normal mode frequencies of clamped nuclei (see Table S1 in the supplementary material) which are added to the total energies at B3LYP/pc-2 and the TC-DFT levels. A detailed inspection of Table S1 in the supplementary material reveals that the computed frequencies at all levels conform well with each other {offering the fact that the vibrational modes of the clamped nuclei are clearly muon-free}. Particularly, intriguing is the fact that the ZPE-corrected total energies computed at B3LYP/pc-2 and B3LYP/pc-2//\emc-1/14s14p14d levels are appreciably closer than those of B3LYP/pc-2 and B3LYP/pc-2//no-\emc/14s14p14d. This indicates that the reliable calculation of ZPEs may be done at the TC-DFT(ee+e$\upmu$) level meeting the second requirement of a self-sufficient computational method. Interestingly, when compared with the difference in the total energies computed at B3LYP/pc-2//\emc-1/14s14p14d and the Numerov levels (see the last column in Table \ref{tab:3}), the remaining difference of the ZPE-corrected energies at B3LYP/pc-2 and B3LYP/pc-2//\emc-1/14s14p14d levels seems to originate from the performance of \emc-1 functional. This suggest that a better designed electron-muon functional may hopefully reproduce B3LYP/pc-2 energies with the chemical accuracy.

One may conclude that the root of problematic aspects of the TC-DFT(ee) predictions is indeed the overlocalization of the muon density, and after the modification of this issue through the inclusion of a proper electron-muon functional, the resulting TC-DFT(ee+e$\upmu$) is a reliable and accurate framework for ab initio computational study of muonic molecules.

\section{Conclusion and outlook}

In this paper, we introduced the first local electron-muon correlation functional which is capable of delivering reasonable total energies, muonic kinetic energies, and muon densities, as a proof-of-principle within the TC-DFT(ee+e$\upmu$) context. This step may hopefully pave the way for future accurate estimation of the hyperfine coupling constants and other local muonic properties that are of utmost importance in the interpretation of the $\upmu$SR spectrum of muonic molecules. Particularly important in this direction is the reliable prediction of muon or Mu binding site to complex molecules which is the topic of a future computational study.   

Beyond the problem of accuracy, solely from a computational perspective, the TC-DFT(ee+e$\upmu$) formulated within the SAA(e$\upmu$/n) framework is more cost effective than those ab initio methods formulated within the DAA and SAA(e/$\upmu$n) frameworks. This stems from the fact that the muonic KS orbital is derived during the SCF cycles applied to the KS equations, which is practically not more time consuming than the SCF cycles applied to the electronic KS equations. Whereas, in the case of the DAA framework, a separate grid-based numerical solution must be found for the muonic Schr\"{o}dinger equation. On the other hand, the muonic KS orbital derived from the TC-DFT calculations naturally contains the anharmonicity of vibrations through using non-s-type Gaussian basis functions (p, d, …) in the muonic basis set. Whereas, in the case of the nuclear plus muon Schr\"{o}dinger equation, derived within the SAA(e/$\upmu$n) framework, the computation of the anharmonic force constants needs the third and higher-order derivatives of energy with respect to the nuclear and muonic position variables which are computationally demanding.\cite{clabo_systematic_1988,barone_vibrational_2004}

{The current parameterization approach is not restricted to the reference data derived from the SAA(e/$\upmu$n) and DAA frameworks, and in general the parameterization may also be done by the data obtained from the explicitly-correlated electron-muon wavefunctions computed within the SAA(e$\upmu$/n) framework.} Furthermore, the inclusion of the gradient correction terms, as demonstrated recently,\cite{tao_multicomponent_2019} may yield non-local electron-muon functionals that hopefully would be superior to the local functional approximation introduced in the present study.  

The \emc-1 functional may also be used to introduce novel effective non-Coulombic potentials, within the context of the newly proposed effective electronic-only Kohn-Sham equations,\cite{rayka_effective_2018,goli_developing_2018} which may overcome the mentioned problem of overlocalization in this context. Let us stress the fact that also from the computational perspective the single-component effective electronic-only equations are extremely cost effective. Developments in this direction is under consideration in our lab.

Accurate TC-DFT(ee+e$\upmu$) results also offer the opportunity to deduce reliable chemical indices, based on the multi-component quantum theory of atoms in molecules,\cite{goli_atoms_2011,goli_two-component_2012,goli_two-component_2013,goli_toward_2013,goli_two-component_2013-1,goli_topological_2015,goli_extending_2016,gharabaghi_incorporating_2017} through analyzing the multi-component KS wavefunctions, which may provide new insights into detailed bonding modes and interatomic interactions in complex muonic systems as demonstrated previously.\cite{goli_deciphering_2014,goli_where_2015,goli_hidden_2015,goli_muon-substituted_2016} This line of research also will be considered in detail in a future computational study.

\section*{Supplementary material}
The supplementary material contains the harmonic normal mode frequencies of the clamped nuclei computed at both B3LYP/pc-2 and the TC-DFT levels, the reduced masses of the muonic normal mode frequencies in Table \ref{tab:1} at B3LYP/pc-2 level, the grid details of the Numerov calculations, and the optimized geometries of the benchmark systems at B3LYP/pc-2 and the TC-DFT levels. 

\begin{acknowledgments}
M.G. is grateful for the kind hospitality provided by the School of Nano Science at the Institute for Research in Fundamental Sciences (IPM). M.G. and S.S. acknowledge the financial support by Iran National Science Foundation (INSF) (Grant No. 98022935), and also the computational resources provided by the SARMAD Cluster at Shahid Beheshti University. \black{The authors are grateful to the anonymous reviewers for their helpful comments and suggestions on a previous draft of this paper.}
\end{acknowledgments}

\appendix

\section{\label{sec:appendix-A}Kohn-Sham potentials}
The electron and muon KS potentials of the \emc-1 functional for a closed-shell system of electrons are the functional derivatives of $E_{\mathrm{e\upmu c}}$ with respect to electron and muon densities, respectively. Using Eq. \ref{eq:closed_shell_EMC}, they are as follows:
\begin{widetext}
\begin{align} 
\nu_{\mathrm{e}}({\br_{\mathrm{e}}}) &=\frac{\delta E_{\mathrm{e\upmu c}}[\erho,\mrho]} {\delta \erho} \nonumber\\
&= -\frac{2\mrho({\br_{\mathrm{e}}}) - {\big(} \mrho({\br_{\mathrm{e}}}){\big )}^{3/2} }{{\Big (}1 + 4\erho({\br_{\mathrm{e}}}) \mrho({\br_{\mathrm{e}}}) + 2\erho({\br_{\mathrm{e}}}) {\big(}\mrho({\br_{\mathrm{e}}}){\big )}^{3/2} {\Big )}^2}, \nonumber
\end{align}
\begin{align}\label{eq:ks_potentials}
\nu_{\upmu}({\br_{\upmu}}) &= \frac{\delta E_{\mathrm{e\upmu c}}[\erho,\mrho]} {\delta \mrho} \nonumber\\
&= - \frac{2\erho({\br_{\upmu}}) - (3/2)\erho({\br_{\upmu}}) {\big(}\mrho({\br_{\upmu}}){\big )}^{1/2} - 4{\big(}\erho({\br_{\upmu}}){\big )}^{2} {\big(}\mrho({\br_{\upmu}}){\big )}^{3/2}  }{{\Big (}1 + 4\erho({\br_{\upmu}})\mrho({\br_{\upmu}}) + 2\erho({\br_{\upmu}}) {\big(}\mrho({\br_{\upmu}}){\big )}^{3/2}{\Big )}^{2}}.
\end{align}
\end{widetext}
The corresponding KS potentials for electronically open-shell systems are derivable using Eq. \ref{eq:polarized_wigner} in a similar manner.

\section*{Data Availability}
The data that supports the findings of this study are available within the article and its supplementary material.


%

\end{document}


\title{Supplementary Material: Two-component density functional theory for muonic molecules: Inclusion of the electron-\black{positive} muon correlation functional}

\author{Mohammad Goli}
\email{m{\_}goli@ipm.ir}
\affiliation{School of Nano Science, Institute for Research in Fundamental Sciences (IPM), Tehran, 19395-5531, Iran}

\author{Shant Shahbazian}
\email{sh{\_}shahbazian@sbu.ac.ir}
\affiliation{Department of Physics, Shahid Beheshti University, Evin, Tehran, Iran}

\date{\today}

\maketitle

\onecolumngrid

\begin{table}[ht]
  \centering
    \caption{The normal mode frequencies ($\mathrm{cm}^{-1}$) of clamped nuclei calculated at B3LYP/pc-2 and the TC-DFT levels for the benchmark systems.}
    \begin{ruledtabular}
    \begin{tabular}{ L{1cm} L{5.25cm} L{5.25cm} L{5.25cm} }
    
    System  & \multicolumn{1}{c}{{B3LYP/pc-2}}   & \multicolumn{1}{c}{\text{B3LYP/pc-2//\emc-1/14s14p14d}} & \multicolumn{1}{c}{{B3LYP/pc-2//no-\emc/14s14p14d}} \\ \hline
1 & 750.7, 830.4, 1131.7, 1645.8, 3034.8, 3243.2 & 765.8, 834.1, 1128.6, 1642.7, 3046.1, 3253.6 & 789.6, 826.7, 1121.6, 1642.3, 3056.2, 3253.9 \vspace{6pt}\\ 
2 & 583.6, 1119.1, 1232.0, 1378.0, 3420.2, 3516.6 & 519.4, 1199.0, 1237.4, 1449.3, 3441.5, 3541.9 &  539.2, 1184.0, 1231.8, 1421.4, 3433.6, 3517.0 \vspace{6pt}\\ 
3 & 116.4, 508.5, 836.6, 1060.2, 1147.5, 1220.0, 1457.7, 1481.4, 3026.8, 3038.8, 3144.7, 3242.5 &  180.2, 508.6, 840.8, 1059.1, 1142.7, 1219.7, 1455.5, 1482.0, 3027.1, 3072.6, 3141.8, 3243.0 &  277.6, 514.7, 840.6, 1066.0, 1124.5, 1222.4, 1454.0, 1483.4, 3032.7, 3084.2, 3141.3, 3241.8 \vspace{6pt}\\ 
4 & 827.8, 1053.9, 1177.2, 1510.6, 2952.6, 3073.4 &  1027.7, 1057.0, 1173.6, 1506.2, 2944.3, 2981.8 & 1036.0, 1104.8, 1179.1, 1508.9, 2939.8, 2987.8 \vspace{6pt}\\ 
5 & 495.1, 1155.2, 1201.3, 1482.2, 3134.0, 3273.2 &  489.0, 1165.7, 1206.2, 1488.4, 3128.0, 3283.2 &  511.5, 1167.8, 1205.3, 1487.3, 3122.2, 3274.9 \vspace{6pt}\\ 
6 & 163.3, 509.9, 775.7, 818.3, 990.4, 1062.3, 1130.5, 1343.4, 1659.3, 2859.7, 3506.3, 3586.4 & 263.4, 526.6, 781.3, 857.4, 994.5, 1083.8, 1134.2, 1355.5, 1662.6, 2899.8, 3525.9, 3607.0 &  202.5, 517.8, 806.1, 867.5, 1009.4, 1110.4, 1163.5, 1376.8, 1662.0, 2919.1, 3523.6, 3605.7 \vspace{6pt}\\
7 & 195.8, 506.4, 663.2, 789.2, 1036.6, 1175.4, 1226.0, 1377.1, 1629.7, 3113.0, 3373.3, 3599.4 & 189.7, 510.8, 665.9, 792.0, 1037.8, 1176.3, 1236.5, 1374.1, 1623.8, 3132.6, 3394.1, 3575.0 & 259.0, 517.8, 662.8, 795.1, 1040.5, 1165.8, 1235.7, 1368.4, 1628.4, 3127.6, 3394.9, 3569.5 \vspace{6pt}\\
8 & 1060.2, 1718.7, 2970.9 & 1059.8, 1717.4, 2973.7 & 1039.5, 1717.6, 2984.7 \vspace{6pt}\\ 
9 & 1055.5, 1788.3, 3020.7 & 1071.0, 1799.7, 3001.3 & 1064.5, 1798.0, 3005.5 \vspace{6pt}\\ 
10 & 258.8, 1001.7, 1042.1, 1128.7, 1350.6, 1474.6, 2935.5, 3067.4, 3383.8 &  358.5, 1023.2, 1041.3, 1123.6, 1347.8, 1483.6, 2977.9, 3066.0, 3406.9 &  419.3, 1021.4, 1045.8, 1131.2, 1346.5, 1490.1, 2988.0, 3065.5, 3413.7 \vspace{6pt}\\ 
11 & 496.4, 590.1, 930.2, 1148.9, 1239.7, 1483.0, 3149.3, 3260.2, 3590.9 &  462.3, 566.2, 975.8, 1229.6, 1266.0, 1485.4, 3153.1, 3266.7, 3613.5 & 462.2, 582.2, 964.4, 1222.6, 1263.0, 1485.6, 3147.5, 3259.2, 3587.5 \vspace{6pt}\\ 
12 & 1344.7, 1442.8, 1444.3, 3056.9, 3112.7, 3113.6 &  1323.4, 1444.4, 1449.2, 3045.4, 3112.1, 3129.9 & 1308.3, 1445.9, 1448.2, 3047.4, 3119.1, 3136.6 \vspace{6pt}\\ 
13 & 698.1, 698.1, 2135.2, 3469.4 &  718.1, 718.3, 2137.2, 3486.8 &  714.1, 714.1, 2131.9, 3486.8 \vspace{6pt}\\ 
14 & 919.2, 984.4, 1029.9, 1286.8, 1431.6, 1667.3, 3134.1, 3167.3, 3215.1 & 940.8, 990.5, 1034.8, 1284.3, 1433.0, 1663.9, 3141.4, 3150.0, 3224.9 &  946.4, 985.5, 1029.7, 1278.6, 1433.6, 1663.9, 3127.5, 3174.1, 3211.8 \vspace{6pt}\\ 
15 & 335.5, 853.3, 913.0, 1015.9, 1241.4, 1283.5, 1423.8, 1471.0, 1504.1, 1505.6, 3028.0, 3043.8, 3060.4, 3086.6, 3092.6 &  414.9, 868.8, 923.2, 1021.6, 1242.8, 1271.5, 1424.9, 1467.0, 1504.6, 1508.1, 3030.0, 3039.9, 3070.4, 3093.4, 3099.5 &  324.1, 858.1, 938.8, 1023.2, 1241.4, 1266.3, 1424.0, 1469.4, 1504.3, 1507.0, 3028.8, 3043.9, 3072.6, 3094.1, 3098.7 \vspace{6pt}\\ 
16 & 415.9, 432.0, 625.4, 627.8, 712.3, 742.6, 870.0, 922.1, 998.9, 1017.2, 1017.8, 1036.0, 1061.7, 1086.0, 1183.3, 1196.0, 1307.6, 1340.3, 1460.7, 1515.4, 1620.0, 1633.2, 3162.0, 3168.6, 3177.0, 3183.9, 3191.9 & 414.5, 436.9, 626.3, 627.9, 716.7, 749.7, 874.6, 937.0, 1006.0, 1017.2, 1021.9, 1035.0, 1063.2, 1087.3, 1181.5, 1198.6, 1308.2, 1344.0, 1462.6, 1515.5, 1617.8, 1630.9, 3152.1, 3169.1, 3177.9, 3191.4, 3205.0 &  413.9, 431.7, 625.3, 628.8, 720.7, 744.1, 870.5, 932.0, 999.2, 1016.5, 1018.9, 1034.8, 1070.4, 1085.0, 1181.7, 1198.6, 1310.3, 1343.2, 1465.4, 1510.7, 1618.4, 1628.9, 3151.6, 3168.6, 3177.5, 3191.0, 3204.6 \vspace{6pt}\\ 
17 & 2289.5 & 2299.1 &  2287.7 \vspace{6pt}\\ 
18 & 1049.0, 1170.1, 1171.0, 1474.4, 1493.1, 1503.9, 2991.2, 3038.3, 3110.0 &  1056.3, 1176.2, 1185.7, 1484.6, 1496.7, 1507.4, 2986.8, 3034.6, 3123.5 &  1049.5, 1173.2, 1183.8, 1479.5, 1493.4, 1506.2, 2980.7, 3027.1, 3116.5 \vspace{6pt}\\ 
19 & 339.7, 966.0, 1041.8, 1131.8, 1265.6, 1456.6, 1499.7, 1513.2, 2968.0, 3057.6, 3093.3, 3542.5 &  309.6, 978.0, 1051.2, 1143.8, 1284.4, 1462.7, 1499.9, 1518.9, 2972.2, 3053.0, 3096.0, 3552.5 &  343.6, 990.8, 1045.2, 1142.9, 1292.1, 1458.0, 1499.7, 1516.8, 2969.8, 3050.4, 3091.5, 3539.9 \vspace{6pt}\\
20 & 1105.0, 1190.5, 1493.1, 1707.1, 3011.2, 3112.5 &  1111.4, 1207.8, 1503.2, 1705.2, 2986.1, 3106.6 &  1106.3, 1206.5, 1501.1, 1703.5, 2988.0, 3102.7
    \end{tabular}%
    \end{ruledtabular}
  \label{tab:s1}%
\end{table}%

\begin{table}[ht]
  \centering
  \caption{The corresponding reduced masses (amu) of the muonic normal mode frequencies, tabulated in Table 1, calculated within the harmonic oscillator approximation at B3LYP/pc-2 level. The on- and off-axis labels refer to the stretching and bending modes, respectively.}
    \begin{ruledtabular}
    \begin{tabular}{l d{1.5}d{1.5}d{1.5}}
     System & \multicolumn{1}{c}{On-axis}    & \multicolumn{1}{c}{Off-axis}   & \multicolumn{1}{c}{Off-axis}   \\ \hline
    1     & 0.11443 & 0.12227 & 0.12454 \\
    2     & 0.11435 & 0.11757 & 0.16015 \\
    3     & 0.11438 & 0.11937 & 0.14167 \\
    4     & 0.11434 & 0.11856 & 0.14993 \\
    5     & 0.11426 & 0.11751 & 0.13236 \\
    6     & 0.11431 & 0.12749 & 0.14494 \\
    7     & 0.11424 & 0.11721 & 0.12908 \\
    8     & 0.11437 & 0.12289 & 0.13447 \\
    9     & 0.11433 & 0.12319 & 0.12647 \\
    10    & 0.11438 & 0.12100 & 0.13924 \\
    11    & 0.11434 & 0.12035 & 0.15413 \\
    12    & 0.11442 & 0.13131 & 0.13161 \\
    13    & 0.11465 & 0.11761 & 0.11761 \\
    14    & 0.11446 & 0.11902 & 0.12077 \\
    15    & 0.11442 & 0.11799 & 0.12729 \\
    16    & 0.11445 & 0.11558 & 0.11687 \\
    17    & 0.11470 & 0.11824 & 0.11824 \\
    18    & 0.11427 & 0.11630 & 0.11747 \\
    19    & 0.11433 & 0.12043 & 0.13597 \\
    20    & 0.11434 & 0.11737 & 0.12096 \\
    \end{tabular}%
    \end{ruledtabular}
  \label{tab:s2}%
\end{table}%

\begin{table}[ht]
  \caption{The 3D grid specification used in the Numerov calculations of the one-particle muonic Schr\"{o}dinger equation. $N_p$, $D$ and $\delta$ represent the numbers of the grid points in each perpendicular direction, the length of each 3D cube edge (\r{A}) and the grid spacing (\r{A}), respectively. The 3D cube is centered at the position of the equilibrium geometry of surrogate clamped hydrogen atom at B3LYP/pc-2 level.}
    \begin{ruledtabular}
    \begin{tabular}{lccc}
     System     & $N_{p}$\footnotemark[1] & $D$\footnotemark[1] & $\delta$  \\ \hline
    1     & 48$\times$48$\times$24 & 2.8$\times$2.8$\times$1.4 & 0.058333 \\
    2     & 48$\times$48$\times$24 & 2.8$\times$2.8$\times$1.4 & 0.058333 \\
    3     & 48$\times$48$\times$24 & 2.8$\times$2.8$\times$1.4 & 0.058333 \\
    4     & 48$\times$48$\times$24 & 2.8$\times$2.8$\times$1.4 & 0.058333 \\
    5     & 48$\times$48$\times$24 & 2.8$\times$2.8$\times$1.4 & 0.058333 \\
    6     & 48$\times$48$\times$24 & 2.8$\times$2.8$\times$1.4 & 0.058333 \\
    7     & 48$\times$48$\times$24 & 2.8$\times$2.8$\times$1.4 & 0.058333 \\
    8     & 48$\times$48$\times$24 & 2.8$\times$2.8$\times$1.4 & 0.058333 \\
    9     & 48$\times$48$\times$24 & 2.8$\times$2.8$\times$1.4 & 0.058333 \\
    10    & 48$\times$48$\times$24 & 2.8$\times$2.8$\times$1.4 & 0.058333 \\
    11    & 48$\times$48$\times$24 & 2.8$\times$2.8$\times$1.4 & 0.058333 \\
    12    & 48$\times$48$\times$24 & 2.8$\times$2.8$\times$1.4 & 0.058333 \\
    13    & 48$\times$48$\times$24 & 2.8$\times$2.8$\times$1.4 & 0.058333 \\
    14    & 48$\times$48$\times$24 & 2.8$\times$2.8$\times$1.4 & 0.058333 \\
    15    & 48$\times$48$\times$24 & 2.8$\times$2.8$\times$1.4 & 0.058333 \\
    16    & 48$\times$48$\times$24 & 2.8$\times$2.8$\times$1.4 & 0.058333 \\
    17    & 48$\times$48$\times$24 & 2.8$\times$2.8$\times$1.4 & 0.058333 \\
    18    & 48$\times$48$\times$24 & 2.8$\times$2.8$\times$1.4 & 0.058333 \\
    19    & 48$\times$48$\times$24 & 2.8$\times$2.8$\times$1.4 & 0.058333 \\
    20    & 48$\times$48$\times$24 & 2.8$\times$2.8$\times$1.4 & 0.058333 \\
    \end{tabular}%
    \end{ruledtabular}
    \footnotetext[1]{off-axis$\times$off-axis$\times$on-axis. The on- and off-axis directions indicate the parallel and perpendicular directions between the grid edge and the vector linking the clamped hydrogen center to its binding site, respectively.}
  \label{tab:s3}%
\end{table}%

\begin{table*}[ht]
  \caption{The optimized geometries (\r{A}) obtained at B3LYP/pc-2 and TC-DFT levels of theory for \textbf{1}.}
  \begin{ruledtabular}
    \begin{tabular}{l d{1.6}d{1.6}d{1.6} d{1.6}d{1.6}d{1.6} d{1.6}d{1.6}d{1.6}}%
         & \multicolumn{3}{c}{B3LYP/pc-2} & \multicolumn{3}{c}{B3LYP/pc-2//\emc-1/14s14p14d} & \multicolumn{3}{c}{{B3LYP/pc-2//no-\emc/14s14p14d}} \\ 
    \cmidrule{2-4} \cmidrule{5-7} \cmidrule{8-10}
    Atom  & \multicolumn{1}{c}{\text{x}} & \multicolumn{1}{c}{\text{y}} & \multicolumn{1}{c}{\text{z}} &
     \multicolumn{1}{c}{\text{x}} & \multicolumn{1}{c}{\text{y}} & \multicolumn{1}{c}{\text{z}} &
      \multicolumn{1}{c}{\text{x}} & \multicolumn{1}{c}{\text{y}} & \multicolumn{1}{c}{\text{z}} \\ \hline
    C     & 0.00000 & 0.00000 & 0.00000 & 0.00000 & 0.00000 & 0.00000 & 0.00000 & 0.00000 & 0.00000 \\
    H     & -1.40853 & 0.00000 & -1.72754 & 1.10212 & 0.00000 & -0.69167 & 1.10081 & 0.00000 & -0.69051 \\
    C     & -1.10003 & 0.00000 & -0.69476 & 1.41367 & 0.00000 & -1.72407 & 1.42050 & 0.00000 & -1.72038 \\
    H     & 0.98266 & 0.00000 & -0.47523 & -0.97656 & 0.00000 & -0.48911 & -0.98223 & 0.00000 & -0.47744 \\
    H\footnotemark[1]    & 0.00000 & 0.00000 & 1.08589 & 0.00000 & 0.00000 & 0.87936 & 0.00000 & 0.00000 & 1.13852 \\
    \end{tabular}%
    \end{ruledtabular}
    \footnotetext[1]{This is the center of the muonic basis set for the TC-DFT optimized geometries. }
  \label{tab:s4}%
\end{table*}%

\begin{table*}[ht]
  \caption{The optimized geometries (\r{A}) obtained at B3LYP/pc-2 and TC-DFT levels of theory for \textbf{2}.}
  \begin{ruledtabular}
    \begin{tabular}{l d{1.6}d{1.6}d{1.6} d{1.6}d{1.6}d{1.6} d{1.6}d{1.6}d{1.6}}%
         & \multicolumn{3}{c}{B3LYP/pc-2} & \multicolumn{3}{c}{B3LYP/pc-2//\emc-1/14s14p14d} & \multicolumn{3}{c}{{B3LYP/pc-2//no-\emc/14s14p14d}} \\
    \cmidrule{2-4} \cmidrule{5-7} \cmidrule{8-10}
    Atom  &  \multicolumn{1}{c}{\text{x}} & \multicolumn{1}{c}{\text{y}} & \multicolumn{1}{c}{\text{z}} &  \multicolumn{1}{c}{\text{x}} & \multicolumn{1}{c}{\text{y}} & \multicolumn{1}{c}{\text{z}} &  \multicolumn{1}{c}{\text{x}} & \multicolumn{1}{c}{\text{y}} & \multicolumn{1}{c}{\text{z}} \\ \hline
    N     & -1.01730 & -0.69480 & -0.54782 & 1.20835 & 0.00000 & -0.60037 & 1.22264 & 0.00000 & -0.57018 \\
    H     & -0.79425 & -0.78993 & -1.53979 & 1.03512 & 0.16331 & -1.59329 & 1.07678 & 0.20291 & -1.56069 \\
    N     & 0.00000 & 0.00000 & 0.00000 & 0.00000 & 0.00000 & 0.00000 & 0.00000 & 0.00000 & 0.00000 \\
    H     & 0.91830 & 0.00000 & -0.42524 & -0.81782 & 0.37592 & -0.45828 & -0.77678 & 0.47170 & -0.44502 \\
    H\footnotemark[1]    & 0.00000 & 0.00000 & 1.00742 & 0.00000 & 0.00000 & 0.75840 & 0.00000 & 0.00000 & 1.04190 \\
    \end{tabular}%
    \end{ruledtabular}
    \footnotetext[1]{This is the center of the muonic basis set for the TC-DFT optimized geometries. }
  \label{tab:s5}%
\end{table*}%

\begin{table*}[ht]
  \caption{The optimized geometries (\r{A}) obtained at B3LYP/pc-2 and TC-DFT levels of theory for \textbf{3}.}
  \begin{ruledtabular}
    \begin{tabular}{l d{1.6}d{1.6}d{1.6} d{1.6}d{1.6}d{1.6} d{1.6}d{1.6}d{1.6}}%
        & \multicolumn{3}{c}{B3LYP/pc-2} & \multicolumn{3}{c}{B3LYP/pc-2//\emc-1/14s14p14d} & \multicolumn{3}{c}{{B3LYP/pc-2//no-\emc/14s14p14d}} \\
    \cmidrule{2-4} \cmidrule{5-7} \cmidrule{8-10}
    Atom  &  \multicolumn{1}{c}{\text{x}} & \multicolumn{1}{c}{\text{y}} & \multicolumn{1}{c}{\text{z}} &  \multicolumn{1}{c}{\text{x}} & \multicolumn{1}{c}{\text{y}} & \multicolumn{1}{c}{\text{z}} &  \multicolumn{1}{c}{\text{x}} & \multicolumn{1}{c}{\text{y}} & \multicolumn{1}{c}{\text{z}} \\ \hline
    C     & 0.00000 & 0.00000 & 0.00000 & 0.00000 & 0.00000 & 0.00000 & 0.00000 & 0.00000 & 0.00000 \\
    H     & -0.56171 & -0.88458 & -0.30681 & -0.56662 & -0.88391 & -0.29913 & -0.56505 & -0.88566 & -0.29563 \\
    H     & -0.56171 & 0.88458 & -0.30681 & -0.56662 & 0.88390 & -0.29915 & -0.56505 & 0.88566 & -0.29563 \\
    C     & 1.37671 & 0.00000 & -0.55259 & 1.36715 & 0.00000 & -0.57420 & 1.37020 & 0.00000 & -0.55318 \\
    H     & 1.91637 & 0.92414 & -0.70258 & 1.90561 & 0.92403 & -0.72861 & 1.91201 & 0.92390 & -0.69672 \\
    H     & 1.91635 & -0.92414 & -0.70263 & 1.90560 & -0.92403 & -0.72863 & 1.91200 & -0.92390 & -0.69675 \\
    H\footnotemark[1]    & 0.00000 & 0.00000 & 1.10032 & 0.00000 & 0.00000 & 0.90485 & 0.00000 & 0.00000 & 1.15695 \\

    \end{tabular}%
    \end{ruledtabular}
    \footnotetext[1]{This is the center of the muonic basis set for the TC-DFT optimized geometries. }
  \label{tab:s6}%
\end{table*}%

\begin{table*}[ht]
  \caption{The optimized geometries (\r{A}) obtained at B3LYP/pc-2 and TC-DFT levels of theory for \textbf{4}.}
  \begin{ruledtabular}
    \begin{tabular}{l d{1.6}d{1.6}d{1.6} d{1.6}d{1.6}d{1.6} d{1.6}d{1.6}d{1.6}}%
        & \multicolumn{3}{c}{B3LYP/pc-2} & \multicolumn{3}{c}{B3LYP/pc-2//\emc-1/14s14p14d} & \multicolumn{3}{c}{{B3LYP/pc-2//no-\emc/14s14p14d}} \\
    \cmidrule{2-4} \cmidrule{5-7} \cmidrule{8-10}
    Atom  &  \multicolumn{1}{c}{\text{x}} & \multicolumn{1}{c}{\text{y}} & \multicolumn{1}{c}{\text{z}} &  \multicolumn{1}{c}{\text{x}} & \multicolumn{1}{c}{\text{y}} & \multicolumn{1}{c}{\text{z}} &  \multicolumn{1}{c}{\text{x}} & \multicolumn{1}{c}{\text{y}} & \multicolumn{1}{c}{\text{z}} \\ \hline
    O     & 1.31435 & 0.00000 & -0.36175 & 1.30023 & 0.00000 & -0.39788 & 1.30185 & 0.00000 & -0.34976 \\
    C     & 0.00000 & 0.00000 & 0.00000 & 0.00000 & 0.00000 & 0.00000 & 0.00000 & 0.00000 & 0.00000 \\
    H     & -0.53508 & 0.90813 & -0.30733 & -0.54640 & 0.90796 & -0.28875 & -0.54464 & 0.91115 & -0.28408 \\
    H     & -0.53509 & -0.90813 & -0.30734 & -0.54642 & -0.90794 & -0.28878 & -0.54458 & -0.91119 & -0.28409 \\
    H\footnotemark[1]    & 0.00000 & 0.00000 & 1.10735 & 0.00000 & 0.00000 & 0.91738 & 0.00000 & 0.00000 & 1.17123 \\

    \end{tabular}%
    \end{ruledtabular}
    \footnotetext[1]{This is the center of the muonic basis set for the TC-DFT optimized geometries. }
  \label{tab:s7}%
\end{table*}%

\begin{table*}[ht]
  \caption{The optimized geometries (\r{A}) obtained at B3LYP/pc-2 and TC-DFT levels of theory for \textbf{5}.}
  \begin{ruledtabular}
    \begin{tabular}{l d{1.6}d{1.6}d{1.6} d{1.6}d{1.6}d{1.6} d{1.6}d{1.6}d{1.6}}%
        & \multicolumn{3}{c}{B3LYP/pc-2} & \multicolumn{3}{c}{B3LYP/pc-2//\emc-1/14s14p14d} & \multicolumn{3}{c}{{B3LYP/pc-2//no-\emc/14s14p14d}} \\
    \cmidrule{2-4} \cmidrule{5-7} \cmidrule{8-10}
    Atom  &  \multicolumn{1}{c}{\text{x}} & \multicolumn{1}{c}{\text{y}} & \multicolumn{1}{c}{\text{z}} &  \multicolumn{1}{c}{\text{x}} & \multicolumn{1}{c}{\text{y}} & \multicolumn{1}{c}{\text{z}} &  \multicolumn{1}{c}{\text{x}} & \multicolumn{1}{c}{\text{y}} & \multicolumn{1}{c}{\text{z}} \\ \hline
    C     & 1.28259 & 0.00000 & -0.46672 & 1.27828 & 0.00000 & -0.47393 & 1.27068 & 0.00000 & -0.49217 \\
    H     & 2.07944 & 0.33172 & 0.18417 & 2.08407 & 0.31705 & 0.17377 & 2.08480 & 0.34340 & 0.13190 \\
    H     & 1.35669 & 0.07631 & -1.53895 & 1.35390 & 0.06864 & -1.54679 & 1.32609 & 0.05756 & -1.56759 \\
    O     & 0.00000 & 0.00000 & 0.00000 & 0.00000 & 0.00000 & 0.00000 & 0.00000 & 0.00000 & 0.00000 \\
    H\footnotemark[1]    & 0.00000 & 0.00000 & 0.96098 & 0.00000 & 0.00000 & 0.67373 & 0.00000 & 0.00000 & 0.98381 \\

    \end{tabular}%
    \end{ruledtabular}
    \footnotetext[1]{This is the center of the muonic basis set for the TC-DFT optimized geometries. }
  \label{tab:s8}%
\end{table*}%

\begin{table*}[ht]
  \caption{The optimized geometries (\r{A}) obtained at B3LYP/pc-2 and TC-DFT levels of theory for \textbf{6}.}
  \begin{ruledtabular}
    \begin{tabular}{l d{1.6}d{1.6}d{1.6} d{1.6}d{1.6}d{1.6} d{1.6}d{1.6}d{1.6}}%
        & \multicolumn{3}{c}{B3LYP/pc-2} & \multicolumn{3}{c}{B3LYP/pc-2//\emc-1/14s14p14d} & \multicolumn{3}{c}{{B3LYP/pc-2//no-\emc/14s14p14d}} \\
    \cmidrule{2-4} \cmidrule{5-7} \cmidrule{8-10}
    Atom  &  \multicolumn{1}{c}{\text{x}} & \multicolumn{1}{c}{\text{y}} & \multicolumn{1}{c}{\text{z}} &  \multicolumn{1}{c}{\text{x}} & \multicolumn{1}{c}{\text{y}} & \multicolumn{1}{c}{\text{z}} &  \multicolumn{1}{c}{\text{x}} & \multicolumn{1}{c}{\text{y}} & \multicolumn{1}{c}{\text{z}} \\ \hline
    N     & -0.61571 & -1.20997 & -0.48049 & 1.35850 & 0.00000 & -0.48474 & 1.36109 & 0.00000 & -0.47511 \\
    H     & -1.58710 & -1.26426 & -0.20142 & 1.82512 & -0.87209 & -0.27001 & 1.79354 & -0.90520 & -0.33788 \\
    H     & -0.57694 & -1.26626 & -1.49025 & 1.38906 & 0.13136 & -1.48806 & 1.40719 & 0.22418 & -1.46124 \\
    O     & -0.50250 & 1.20524 & -0.39135 & -0.84619 & -0.98820 & -0.38965 & -0.81649 & -1.02372 & -0.29962 \\
    C     & 0.00000 & 0.00000 & 0.00000 & 0.00000 & 0.00000 & 0.00000 & 0.00000 & 0.00000 & 0.00000 \\
    H     & 1.07172 & 0.00000 & -0.26802 & -0.46984 & 0.97131 & -0.23282 & -0.47390 & 0.96852 & -0.22955 \\
    H\footnotemark[1]    & 0.00000 & 0.00000 & 1.10469 & 0.00000 & 0.00000 & 0.92243 & 0.00000 & 0.00000 & 1.17774 \\

    \end{tabular}%
    \end{ruledtabular}
    \footnotetext[1]{This is the center of the muonic basis set for the TC-DFT optimized geometries. }
  \label{tab:s9}%
\end{table*}%

\begin{table*}[ht]
  \caption{The optimized geometries (\r{A}) obtained at B3LYP/pc-2 and TC-DFT levels of theory for \textbf{7}.}
  \begin{ruledtabular}
    \begin{tabular}{l d{1.6}d{1.6}d{1.6} d{1.6}d{1.6}d{1.6} d{1.6}d{1.6}d{1.6}}%
        & \multicolumn{3}{c}{B3LYP/pc-2} & \multicolumn{3}{c}{B3LYP/pc-2//\emc-1/14s14p14d} & \multicolumn{3}{c}{{B3LYP/pc-2//no-\emc/14s14p14d}} \\
    \cmidrule{2-4} \cmidrule{5-7} \cmidrule{8-10}
    Atom  &  \multicolumn{1}{c}{\text{x}} & \multicolumn{1}{c}{\text{y}} & \multicolumn{1}{c}{\text{z}} &  \multicolumn{1}{c}{\text{x}} & \multicolumn{1}{c}{\text{y}} & \multicolumn{1}{c}{\text{z}} &  \multicolumn{1}{c}{\text{x}} & \multicolumn{1}{c}{\text{y}} & \multicolumn{1}{c}{\text{z}} \\ \hline
    N     & 2.18736 & -0.69659 & 0.42736 & 2.19054 & -0.68460 & 0.43665 & 2.19635 & -0.65156 & 0.45769 \\
    H     & 2.04386 & -1.70539 & 0.46813 & 2.05645 & -1.69481 & 0.47887 & 2.11559 & -1.66823 & 0.47614 \\
    H     & 3.15820 & -0.49747 & 0.23881 & 3.15982 & -0.47937 & 0.24561 & 3.15981 & -0.39570 & 0.29961 \\
    C     & 1.30486 & 0.00000 & -0.41077 & 1.30373 & 0.00000 & -0.40907 & 1.30050 & 0.00000 & -0.40942 \\
    H     & 1.38611 & -0.00361 & -1.49358 & 1.39435 & -0.00847 & -1.49122 & 1.39039 & -0.04223 & -1.49137 \\
    O     & 0.00000 & 0.00000 & 0.00000 & 0.00000 & 0.00000 & 0.00000 & 0.00000 & 0.00000 & 0.00000 \\
    H\footnotemark[1]    & 0.00000 & 0.00000 & 0.96508 & 0.00000 & 0.00000 & 0.68298 & 0.00000 & 0.00000 & 0.98941 \\

    \end{tabular}%
    \end{ruledtabular}
    \footnotetext[1]{This is the center of the muonic basis set for the TC-DFT optimized geometries. }
  \label{tab:s10}%
\end{table*}%

\begin{table*}[ht]
  \caption{The optimized geometries (\r{A}) obtained at B3LYP/pc-2 and TC-DFT levels of theory for \textbf{8}.}
  \begin{ruledtabular}
    \begin{tabular}{l d{1.6}d{1.6}d{1.6} d{1.6}d{1.6}d{1.6} d{1.6}d{1.6}d{1.6}}%
        & \multicolumn{3}{c}{B3LYP/pc-2} & \multicolumn{3}{c}{B3LYP/pc-2//\emc-1/14s14p14d} & \multicolumn{3}{c}{{B3LYP/pc-2//no-\emc/14s14p14d}} \\
    \cmidrule{2-4} \cmidrule{5-7} \cmidrule{8-10}
    Atom  &  \multicolumn{1}{c}{\text{x}} & \multicolumn{1}{c}{\text{y}} & \multicolumn{1}{c}{\text{z}} &  \multicolumn{1}{c}{\text{x}} & \multicolumn{1}{c}{\text{y}} & \multicolumn{1}{c}{\text{z}} &  \multicolumn{1}{c}{\text{x}} & \multicolumn{1}{c}{\text{y}} & \multicolumn{1}{c}{\text{z}} \\ \hline
    C     & 0.00000 & 0.00000 & 0.00000 & 0.00000 & 0.00000 & 0.00000 & 0.00000 & 0.00000 & 0.00000 \\
    H     & 0.97904 & 0.00000 & -0.49558 & -0.98544 & 0.00000 & -0.48581 & -0.98132 & 0.00052 & -0.49330 \\
    N     & -1.05447 & 0.00000 & -0.64804 & 1.04119 & 0.00000 & -0.66854 & 1.05399 & 0.00000 & -0.64236 \\
    H\footnotemark[1]    & 0.00000 & 0.00000 & 1.09735 & 0.00000 & 0.00000 & 0.89930 & 0.00000 & 0.00000 & 1.15599 \\

    \end{tabular}%
    \end{ruledtabular}
    \footnotetext[1]{This is the center of the muonic basis set for the TC-DFT optimized geometries. }
  \label{tab:s11}%
\end{table*}%

\begin{table*}[ht]
  \caption{The optimized geometries (\r{A}) obtained at B3LYP/pc-2 and TC-DFT levels of theory for \textbf{9}.}
  \begin{ruledtabular}
    \begin{tabular}{l d{1.6}d{1.6}d{1.6} d{1.6}d{1.6}d{1.6} d{1.6}d{1.6}d{1.6}}%
        & \multicolumn{3}{c}{B3LYP/pc-2} & \multicolumn{3}{c}{B3LYP/pc-2//\emc-1/14s14p14d} & \multicolumn{3}{c}{{B3LYP/pc-2//no-\emc/14s14p14d}} \\
    \cmidrule{2-4} \cmidrule{5-7} \cmidrule{8-10}
    Atom  &  \multicolumn{1}{c}{\text{x}} & \multicolumn{1}{c}{\text{y}} & \multicolumn{1}{c}{\text{z}} &  \multicolumn{1}{c}{\text{x}} & \multicolumn{1}{c}{\text{y}} & \multicolumn{1}{c}{\text{z}} &  \multicolumn{1}{c}{\text{x}} & \multicolumn{1}{c}{\text{y}} & \multicolumn{1}{c}{\text{z}} \\ \hline
    C     & 1.06872 & 0.00000 & -0.60375 & 1.06985 & 0.00000 & -0.60077 & 1.06265 & 0.00000 & -0.61196 \\
    H     & 1.17806 & 0.00000 & -1.69329 & 1.18255 & 0.00000 & -1.69199 & 1.16341 & -0.00004 & -1.70420 \\
    N     & 0.00000 & 0.00000 & 0.00000 & 0.00000 & 0.00000 & 0.00000 & 0.00000 & 0.00000 & 0.00000 \\
    H\footnotemark[1]    & 0.00000 & 0.00000 & 1.01944 & 0.00000 & 0.00000 & 0.77282 & 0.00000 & 0.00000 & 1.06013 \\

    \end{tabular}%
    \end{ruledtabular}
    \footnotetext[1]{This is the center of the muonic basis set for the TC-DFT optimized geometries. }
  \label{tab:s12}%
\end{table*}%

\begin{table*}[ht]
  
  \caption{The optimized geometries (\r{A}) obtained at B3LYP/pc-2 and TC-DFT levels of theory for \textbf{10}.}
  \begin{ruledtabular}
    \begin{tabular}{l d{1.6}d{1.6}d{1.6} d{1.6}d{1.6}d{1.6} d{1.6}d{1.6}d{1.6}}%
        & \multicolumn{3}{c}{B3LYP/pc-2} & \multicolumn{3}{c}{B3LYP/pc-2//\emc-1/14s14p14d} & \multicolumn{3}{c}{{B3LYP/pc-2//no-\emc/14s14p14d}} \\
    \cmidrule{2-4} \cmidrule{5-7} \cmidrule{8-10}
    Atom  &  \multicolumn{1}{c}{\text{x}} & \multicolumn{1}{c}{\text{y}} & \multicolumn{1}{c}{\text{z}} &  \multicolumn{1}{c}{\text{x}} & \multicolumn{1}{c}{\text{y}} & \multicolumn{1}{c}{\text{z}} &  \multicolumn{1}{c}{\text{x}} & \multicolumn{1}{c}{\text{y}} & \multicolumn{1}{c}{\text{z}} \\ \hline
    N     & 1.33737 & 0.00000 & -0.52680 & 1.33447 & 0.00000 & -0.53032 & 1.34717 & 0.00000 & -0.47718 \\
    H     & 1.80490 & -0.83768 & -0.16644 & 1.74886 & -0.91309 & -0.31997 & 1.72036 & -0.94617 & -0.35767 \\
    C     & 0.00000 & 0.00000 & 0.00000 & 0.00000 & 0.00000 & 0.00000 & 0.00000 & 0.00000 & 0.00000 \\
    H     & -0.55097 & 0.87259 & -0.34882 & -0.54265 & 0.88507 & -0.33092 & -0.52212 & 0.90407 & -0.31311 \\
    H     & -0.54963 & -0.90520 & -0.29681 & -0.57221 & -0.89652 & -0.27354 & -0.58272 & -0.88377 & -0.28814 \\
    H\footnotemark[1]    & 0.00000 & 0.00000 & 1.09980 & 0.00000 & 0.00000 & 0.91009 & 0.00000 & 0.00000 & 1.15946 \\

    \end{tabular}%
    \end{ruledtabular}
    \footnotetext[1]{This is the center of the muonic basis set for the TC-DFT optimized geometries. }
  \label{tab:s13}%
\end{table*}%

\begin{table*}[ht]
  \caption{The optimized geometries (\r{A}) obtained at B3LYP/pc-2 and TC-DFT levels of theory for \textbf{11}.}
  \begin{ruledtabular}
    \begin{tabular}{l d{1.6}d{1.6}d{1.6} d{1.6}d{1.6}d{1.6} d{1.6}d{1.6}d{1.6}}%
        & \multicolumn{3}{c}{B3LYP/pc-2} & \multicolumn{3}{c}{B3LYP/pc-2//\emc-1/14s14p14d} & \multicolumn{3}{c}{{B3LYP/pc-2//no-\emc/14s14p14d}} \\
    \cmidrule{2-4} \cmidrule{5-7} \cmidrule{8-10}
    Atom  &  \multicolumn{1}{c}{\text{x}} & \multicolumn{1}{c}{\text{y}} & \multicolumn{1}{c}{\text{z}} &  \multicolumn{1}{c}{\text{x}} & \multicolumn{1}{c}{\text{y}} & \multicolumn{1}{c}{\text{z}} &  \multicolumn{1}{c}{\text{x}} & \multicolumn{1}{c}{\text{y}} & \multicolumn{1}{c}{\text{z}} \\ \hline
    C     & 1.24970 & 0.00000 & -0.61121 & 1.22232 & 0.00000 & -0.66217 & 1.23814 & 0.00000 & -0.63342 \\
    H     & 2.02287 & 0.58961 & -0.14278 & 2.02900 & 0.55301 & -0.20628 & 2.01769 & 0.60149 & -0.19097 \\
    H     & 1.27035 & -0.13173 & -1.68219 & 1.20415 & -0.10970 & -1.73551 & 1.24482 & -0.15632 & -1.70183 \\
    N     & 0.00000 & 0.00000 & 0.00000 & 0.00000 & 0.00000 & 0.00000 & 0.00000 & 0.00000 & 0.00000 \\
    H     & -0.67507 & -0.64875 & -0.37277 & -0.72524 & -0.57531 & -0.39460 & -0.68415 & -0.63506 & -0.38057 \\
    H\footnotemark[1]    & 0.00000 & 0.00000 & 1.00783 & 0.00000 & 0.00000 & 0.75310 & 0.00000 & 0.00000 & 1.04120 \\

    \end{tabular}%
    \end{ruledtabular}
    \footnotetext[1]{This is the center of the muonic basis set for the TC-DFT optimized geometries. }
  \label{tab:s14}%
\end{table*}%

\begin{table*}[ht]
  \caption{The optimized geometries (\r{A}) obtained at B3LYP/pc-2 and TC-DFT levels of theory for \textbf{12}.}
  \begin{ruledtabular}
    \begin{tabular}{l d{1.6}d{1.6}d{1.6} d{1.6}d{1.6}d{1.6} d{1.6}d{1.6}d{1.6}}%
        & \multicolumn{3}{c}{B3LYP/pc-2} & \multicolumn{3}{c}{B3LYP/pc-2//\emc-1/14s14p14d} & \multicolumn{3}{c}{{B3LYP/pc-2//no-\emc/14s14p14d}} \\
    \cmidrule{2-4} \cmidrule{5-7} \cmidrule{8-10}
    Atom  &  \multicolumn{1}{c}{\text{x}} & \multicolumn{1}{c}{\text{y}} & \multicolumn{1}{c}{\text{z}} &  \multicolumn{1}{c}{\text{x}} & \multicolumn{1}{c}{\text{y}} & \multicolumn{1}{c}{\text{z}} &  \multicolumn{1}{c}{\text{x}} & \multicolumn{1}{c}{\text{y}} & \multicolumn{1}{c}{\text{z}} \\ \hline
    C     & 0.00000 & 0.00000 & 0.00000 & 0.00000 & 0.00000 & 0.00000 & 0.00000 & 0.00000 & 0.00000 \\
    H     & -0.51305 & 0.88863 & -0.36278 & -0.51270 & -0.88774 & -0.36550 & 1.02639 & 0.00000 & -0.36167 \\
    H     & 1.02610 & 0.00000 & -0.36278 & -0.51230 & 0.88743 & -0.36671 & -0.51307 & -0.88897 & -0.36182 \\
    H     & -0.51305 & -0.88863 & -0.36278 & 1.02514 & 0.00000 & -0.36556 & -0.51346 & 0.88872 & -0.36168 \\
    H\footnotemark[1]    & 0.00000 & 0.00000 & 1.08834 & 0.00000 & 0.00000 & 0.88512 & 0.00000 & 0.00000 & 1.13750 \\

    \end{tabular}%
    \end{ruledtabular}
    \footnotetext[1]{This is the center of the muonic basis set for the TC-DFT optimized geometries. }
  \label{tab:s15}%
\end{table*}%

\begin{table*}[ht]
  \caption{The optimized geometries (\r{A}) obtained at B3LYP/pc-2 and TC-DFT levels of theory for \textbf{13}.}
  \begin{ruledtabular}
    \begin{tabular}{l d{1.6}d{1.6}d{1.6} d{1.6}d{1.6}d{1.6} d{1.6}d{1.6}d{1.6}}%
        & \multicolumn{3}{c}{B3LYP/pc-2} & \multicolumn{3}{c}{B3LYP/pc-2//\emc-1/14s14p14d} & \multicolumn{3}{c}{{B3LYP/pc-2//no-\emc/14s14p14d}} \\
    \cmidrule{2-4} \cmidrule{5-7} \cmidrule{8-10}
    Atom  &  \multicolumn{1}{c}{\text{x}} & \multicolumn{1}{c}{\text{y}} & \multicolumn{1}{c}{\text{z}} &  \multicolumn{1}{c}{\text{x}} & \multicolumn{1}{c}{\text{y}} & \multicolumn{1}{c}{\text{z}} &  \multicolumn{1}{c}{\text{x}} & \multicolumn{1}{c}{\text{y}} & \multicolumn{1}{c}{\text{z}} \\ \hline
    C     & 0.00000 & 0.00000 & 0.00000 & 0.00000 & 0.00000 & 0.00000 & 0.00000 & 0.00000 & 0.00000 \\
    C     & 0.00000 & 0.00000 & -1.19636 & 0.00000 & 0.00000 & -1.19749 & 0.00000 & 0.00000 & -1.19833 \\
    H     & 0.00000 & 0.00000 & -2.25848 & 0.00000 & 0.00000 & -2.25998 & 0.00000 & 0.00000 & -2.26068 \\
    H\footnotemark[1]    & 0.00000 & 0.00000 & 1.06212 & 0.00000 & 0.00000 & 0.82646 & 0.00000 & 0.00000 & 1.10634 \\

    \end{tabular}%
    \end{ruledtabular}
    \footnotetext[1]{This is the center of the muonic basis set for the TC-DFT optimized geometries. }
  \label{tab:s16}%
\end{table*}%

\begin{table*}[ht]
  \caption{The optimized geometries (\r{A}) obtained at B3LYP/pc-2 and TC-DFT levels of theory for \textbf{14}.}
  \begin{ruledtabular}
    \begin{tabular}{l d{1.6}d{1.6}d{1.6} d{1.6}d{1.6}d{1.6} d{1.6}d{1.6}d{1.6}}%
        & \multicolumn{3}{c}{B3LYP/pc-2} & \multicolumn{3}{c}{B3LYP/pc-2//\emc-1/14s14p14d} & \multicolumn{3}{c}{{B3LYP/pc-2//no-\emc/14s14p14d}} \\
    \cmidrule{2-4} \cmidrule{5-7} \cmidrule{8-10}
    Atom  &  \multicolumn{1}{c}{\text{x}} & \multicolumn{1}{c}{\text{y}} & \multicolumn{1}{c}{\text{z}} &  \multicolumn{1}{c}{\text{x}} & \multicolumn{1}{c}{\text{y}} & \multicolumn{1}{c}{\text{z}} &  \multicolumn{1}{c}{\text{x}} & \multicolumn{1}{c}{\text{y}} & \multicolumn{1}{c}{\text{z}} \\ \hline
    C     & 0.00000 & 0.00000 & 0.00000 & 0.00000 & 0.00000 & 0.00000 & 0.00000 & 0.00000 & 0.00000 \\
    C     & 1.12636 & 0.00000 & -0.69714 & -1.12271 & 0.00000 & -0.70504 & 1.12249 & 0.00000 & -0.70474 \\
    H     & -0.96903 & 0.00000 & -0.48295 & 0.96543 & 0.00000 & -0.49083 & -0.96758 & 0.00000 & -0.48712 \\
    H     & 1.12636 & 0.00000 & -1.77984 & -1.11711 & 0.00000 & -1.78826 & 1.11598 & 0.00000 & -1.78830 \\
    H     & 2.09539 & 0.00000 & -0.21419 & -2.09624 & 0.00000 & -0.23058 & 2.09675 & 0.00000 & -0.23185 \\
    H\footnotemark[1]    & 0.00000 & 0.00000 & 1.08271 & 0.00000 & 0.00000 & 0.87830 & 0.00000 & 0.00000 & 1.13250 \\

    \end{tabular}%
    \end{ruledtabular}
    \footnotetext[1]{This is the center of the muonic basis set for the TC-DFT optimized geometries. }
  \label{tab:s17}%
\end{table*}%

\begin{table*}[ht]
  \caption{The optimized geometries (\r{A}) obtained at B3LYP/pc-2 and TC-DFT levels of theory for \textbf{15}.}
  \begin{ruledtabular}
    \begin{tabular}{l d{1.6}d{1.6}d{1.6} d{1.6}d{1.6}d{1.6} d{1.6}d{1.6}d{1.6}}%
        & \multicolumn{3}{c}{B3LYP/pc-2} & \multicolumn{3}{c}{B3LYP/pc-2//\emc-1/14s14p14d} & \multicolumn{3}{c}{{B3LYP/pc-2//no-\emc/14s14p14d}} \\
    \cmidrule{2-4} \cmidrule{5-7} \cmidrule{8-10}
    Atom  &  \multicolumn{1}{c}{\text{x}} & \multicolumn{1}{c}{\text{y}} & \multicolumn{1}{c}{\text{z}} &  \multicolumn{1}{c}{\text{x}} & \multicolumn{1}{c}{\text{y}} & \multicolumn{1}{c}{\text{z}} &  \multicolumn{1}{c}{\text{x}} & \multicolumn{1}{c}{\text{y}} & \multicolumn{1}{c}{\text{z}} \\ \hline
    C     & 1.42182 & 0.00000 & -0.55686 & 1.41887 & 0.00000 & -0.56871 & 1.41843 & 0.00000 & -0.56674 \\
    H     & 1.97763 & -0.87993 & -0.22895 & 1.97765 & -0.87984 & -0.24569 & 1.97750 & -0.87975 & -0.24383 \\
    H     & 1.42182 & 0.00000 & -1.64807 & 1.41021 & 0.00001 & -1.66011 & 1.41070 & 0.00000 & -1.65847 \\
    H     & 1.97763 & 0.87993 & -0.22895 & 1.97765 & 0.87984 & -0.24569 & 1.97750 & 0.87975 & -0.24383 \\
    C     & 0.00000 & 0.00000 & 0.00000 & 0.00000 & 0.00000 & 0.00000 & 0.00000 & 0.00000 & 0.00000 \\
    H     & -0.55581 & -0.87993 & -0.32792 & -0.55717 & -0.87856 & -0.32891 & -0.55802 & -0.87974 & -0.32395 \\
    H     & -0.55581 & 0.87993 & -0.32792 & -0.55718 & 0.87856 & -0.32891 & -0.55802 & 0.87974 & -0.32395 \\
    H\footnotemark[1]    & 0.00000 & 0.00000 & 1.09121 & 0.00000 & 0.00000 & 0.89361 & 0.00000 & 0.00000 & 1.14187 \\

    \end{tabular}%
    \end{ruledtabular}
    \footnotetext[1]{This is the center of the muonic basis set for the TC-DFT optimized geometries. }
  \label{tab:s18}%
\end{table*}%

\begin{table*}[ht]
  \caption{The optimized geometries (\r{A}) obtained at B3LYP/pc-2 and TC-DFT levels of theory for \textbf{16}.}
  \begin{ruledtabular}
    \begin{tabular}{l d{1.6}d{1.6}d{1.6} d{1.6}d{1.6}d{1.6} d{1.6}d{1.6}d{1.6}}%
        & \multicolumn{3}{c}{B3LYP/pc-2} & \multicolumn{3}{c}{B3LYP/pc-2//\emc-1/14s14p14d} & \multicolumn{3}{c}{{B3LYP/pc-2//no-\emc/14s14p14d}} \\
    \cmidrule{2-4} \cmidrule{5-7} \cmidrule{8-10}
    Atom  &  \multicolumn{1}{c}{\text{x}} & \multicolumn{1}{c}{\text{y}} & \multicolumn{1}{c}{\text{z}} &  \multicolumn{1}{c}{\text{x}} & \multicolumn{1}{c}{\text{y}} & \multicolumn{1}{c}{\text{z}} &  \multicolumn{1}{c}{\text{x}} & \multicolumn{1}{c}{\text{y}} & \multicolumn{1}{c}{\text{z}} \\ \hline
    C     & 1.20467 & 0.00000 & -2.08656 & -1.20411 & -0.00130 & -2.09205 & 0.03560 & -1.20359 & -2.09086 \\
    C     & 0.00000 & 0.00000 & -2.78207 & 0.00004 & -0.00122 & -2.78787 & 0.00280 & 0.00000 & -2.78682 \\
    C     & -1.20467 & 0.00000 & -2.08656 & 1.20417 & -0.00074 & -2.09200 & -0.03086 & 1.20361 & -2.09093 \\
    C     & -1.20467 & 0.00000 & -0.69552 & 1.20322 & 0.00000 & -0.70082 & -0.03238 & 1.20318 & -0.69943 \\
    C     & 0.00000 & 0.00000 & 0.00000 & 0.00000 & 0.00000 & 0.00000 & 0.00000 & 0.00000 & 0.00000 \\
    C     & 1.20467 & 0.00000 & -0.69552 & -1.20319 & -0.00071 & -0.70086 & 0.03399 & -1.20319 & -0.69937 \\
    H     & 2.14183 & 0.00000 & -2.62762 & -2.14177 & -0.00205 & -2.63228 & 0.06302 & -2.14093 & -2.63109 \\
    H     & 0.00000 & 0.00000 & -3.86421 & 0.00002 & -0.00146 & -3.87015 & 0.00354 & -0.00004 & -3.86909 \\
    H     & -2.14183 & 0.00000 & -2.62762 & 2.14183 & -0.00112 & -2.63225 & -0.05532 & 2.14096 & -2.63128 \\
    H     & -2.14183 & 0.00000 & -0.15445 & 2.14297 & 0.00043 & -0.16365 & -0.05847 & 2.14253 & -0.16245 \\
    H     & 2.14183 & 0.00000 & -0.15445 & -2.14296 & -0.00077 & -0.16372 & 0.05955 & -2.14250 & -0.16231 \\
    H\footnotemark[1]    & 0.00000 & 0.00000 & 1.08214 & 0.00000 & 0.00000 & 0.87631 & 0.00000 & 0.00000 & 1.13277 \\

    \end{tabular}%
    \end{ruledtabular}
    \footnotetext[1]{This is the center of the muonic basis set for the TC-DFT optimized geometries. }
  \label{tab:s19}%
\end{table*}%

\begin{table*}[ht]
  \caption{The optimized geometries (\r{A}) obtained at B3LYP/pc-2 and TC-DFT levels of theory for \textbf{17}.}
  \begin{ruledtabular}
    \begin{tabular}{l d{1.6}d{1.6}d{1.6} d{1.6}d{1.6}d{1.6} d{1.6}d{1.6}d{1.6}}%
        & \multicolumn{3}{c}{B3LYP/pc-2} & \multicolumn{3}{c}{B3LYP/pc-2//\emc-1/14s14p14d} & \multicolumn{3}{c}{{B3LYP/pc-2//no-\emc/14s14p14d}} \\
    \cmidrule{2-4} \cmidrule{5-7} \cmidrule{8-10}
    Atom  &  \multicolumn{1}{c}{\text{x}} & \multicolumn{1}{c}{\text{y}} & \multicolumn{1}{c}{\text{z}} &  \multicolumn{1}{c}{\text{x}} & \multicolumn{1}{c}{\text{y}} & \multicolumn{1}{c}{\text{z}} &  \multicolumn{1}{c}{\text{x}} & \multicolumn{1}{c}{\text{y}} & \multicolumn{1}{c}{\text{z}} \\ \hline
    C     & 0.00000 & 0.00000 & 0.00000 & 0.00000 & 0.00000 & 0.00000 & 0.00000 & 0.00000 & 0.00000 \\
    N     & 0.00000 & 0.00000 & -1.14604 & 0.00000 & 0.00000 & -1.14598 & 0.00000 & 0.00000 & -1.14737 \\
    H\footnotemark[1]    & 0.00000 & 0.00000 & 1.06602 & 0.00000 & 0.00000 & 0.82596 & 0.00000 & 0.00000 & 1.11091 \\

    \end{tabular}%
    \end{ruledtabular}
    \footnotetext[1]{This is the center of the muonic basis set for the TC-DFT optimized geometries. }
  \label{tab:s20}%
\end{table*}%

\begin{table*}[ht]
  \caption{The optimized geometries (\r{A}) obtained at B3LYP/pc-2 and TC-DFT levels of theory for \textbf{18}.}
  \begin{ruledtabular}
    \begin{tabular}{l d{1.6}d{1.6}d{1.6} d{1.6}d{1.6}d{1.6} d{1.6}d{1.6}d{1.6}}%
        & \multicolumn{3}{c}{B3LYP/pc-2} & \multicolumn{3}{c}{B3LYP/pc-2//\emc-1/14s14p14d} & \multicolumn{3}{c}{{B3LYP/pc-2//no-\emc/14s14p14d}} \\
    \cmidrule{2-4} \cmidrule{5-7} \cmidrule{8-10}
    Atom  &  \multicolumn{1}{c}{\text{x}} & \multicolumn{1}{c}{\text{y}} & \multicolumn{1}{c}{\text{z}} &  \multicolumn{1}{c}{\text{x}} & \multicolumn{1}{c}{\text{y}} & \multicolumn{1}{c}{\text{z}} &  \multicolumn{1}{c}{\text{x}} & \multicolumn{1}{c}{\text{y}} & \multicolumn{1}{c}{\text{z}} \\ \hline
    C     & 1.34543 & 0.00000 & -0.46088 & 1.34021 & 0.00000 & -0.46881 & 1.33375 & 0.00000 & -0.48888 \\
    H     & 1.30581 & 0.00000 & -1.54831 & 1.30064 & -0.00034 & -1.55660 & 1.27473 & -0.00001 & -1.57642 \\
    H     & 1.89224 & 0.89053 & -0.13546 & 1.89067 & 0.88998 & -0.14684 & 1.89178 & 0.88958 & -0.17758 \\
    H     & 1.89224 & -0.89053 & -0.13546 & 1.89123 & -0.88936 & -0.14615 & 1.89179 & -0.88957 & -0.17756 \\
    O     & 0.00000 & 0.00000 & 0.00000 & 0.00000 & 0.00000 & 0.00000 & 0.00000 & 0.00000 & 0.00000 \\
    H\footnotemark[1]    & 0.00000 & 0.00000 & 0.95983 & 0.00000 & 0.00000 & 0.66906 & 0.00000 & 0.00000 & 0.98173 \\

    \end{tabular}%
    \end{ruledtabular}
    \footnotetext[1]{This is the center of the muonic basis set for the TC-DFT optimized geometries. }
  \label{tab:s21}%
\end{table*}%

\begin{table*}[ht]
  \caption{The optimized geometries (\r{A}) obtained at B3LYP/pc-2 and TC-DFT levels of theory for \textbf{19}.}
  \begin{ruledtabular}
    \begin{tabular}{l d{1.6}d{1.6}d{1.6} d{1.6}d{1.6}d{1.6} d{1.6}d{1.6}d{1.6}}%
        & \multicolumn{3}{c}{B3LYP/pc-2} & \multicolumn{3}{c}{B3LYP/pc-2//\emc-1/14s14p14d} & \multicolumn{3}{c}{{B3LYP/pc-2//no-\emc/14s14p14d}} \\
    \cmidrule{2-4} \cmidrule{5-7} \cmidrule{8-10}
    Atom  &  \multicolumn{1}{c}{\text{x}} & \multicolumn{1}{c}{\text{y}} & \multicolumn{1}{c}{\text{z}} &  \multicolumn{1}{c}{\text{x}} & \multicolumn{1}{c}{\text{y}} & \multicolumn{1}{c}{\text{z}} &  \multicolumn{1}{c}{\text{x}} & \multicolumn{1}{c}{\text{y}} & \multicolumn{1}{c}{\text{z}} \\ \hline
    C     & 1.36564 & 0.00000 & -0.52483 & 1.34608 & 0.00000 & -0.56940 & 1.35641 & 0.00000 & -0.54868 \\
    H     & 1.33161 & 0.01881 & -1.61454 & 1.28215 & 0.04628 & -1.65732 & 1.30643 & 0.03252 & -1.63832 \\
    H     & 1.87396 & 0.90953 & -0.20333 & 1.87680 & 0.89513 & -0.24275 & 1.87738 & 0.90242 & -0.22634 \\
    H     & 1.98190 & -0.85611 & -0.22131 & 1.96253 & -0.86934 & -0.30614 & 1.97541 & -0.86223 & -0.26788 \\
    N     & 0.00000 & 0.00000 & 0.00000 & 0.00000 & 0.00000 & 0.00000 & 0.00000 & 0.00000 & 0.00000 \\
    H     & -0.50295 & -0.82598 & -0.29706 & -0.53423 & -0.80047 & -0.30757 & -0.51094 & -0.82011 & -0.30001 \\
    H\footnotemark[1]    & 0.00000 & 0.00000 & 1.01165 & 0.00000 & 0.00000 & 0.76081 & 0.00000 & 0.00000 & 1.04397 \\

    \end{tabular}%
    \end{ruledtabular}
    \footnotetext[1]{This is the center of the muonic basis set for the TC-DFT optimized geometries. }
  \label{tab:s22}%
\end{table*}%

\begin{table*}[ht]
  \caption{The optimized geometries (\r{A}) obtained at B3LYP/pc-2 and TC-DFT levels of theory for \textbf{20}.}
  \begin{ruledtabular}
    \begin{tabular}{l d{1.6}d{1.6}d{1.6} d{1.6}d{1.6}d{1.6} d{1.6}d{1.6}d{1.6}}%
        & \multicolumn{3}{c}{B3LYP/pc-2} & \multicolumn{3}{c}{B3LYP/pc-2//\emc-1/14s14p14d} & \multicolumn{3}{c}{{B3LYP/pc-2//no-\emc/14s14p14d}} \\
    \cmidrule{2-4} \cmidrule{5-7} \cmidrule{8-10}
    Atom  &  \multicolumn{1}{c}{\text{x}} & \multicolumn{1}{c}{\text{y}} & \multicolumn{1}{c}{\text{z}} &  \multicolumn{1}{c}{\text{x}} & \multicolumn{1}{c}{\text{y}} & \multicolumn{1}{c}{\text{z}} &  \multicolumn{1}{c}{\text{x}} & \multicolumn{1}{c}{\text{y}} & \multicolumn{1}{c}{\text{z}} \\ \hline
    C     & 1.17648 & 0.00000 & -0.46043 & 1.18262 & 0.00000 & -0.44239 & 1.16992 & 0.00000 & -0.47569 \\
    H     & 1.31950 & 0.00000 & -1.54036 & 1.34708 & 0.00000 & -1.52085 & 1.29951 & 0.00000 & -1.55921 \\
    H     & 2.08580 & 0.00000 & 0.14789 & 2.08662 & 0.00000 & 0.17621 & 2.09349 & 0.00000 & 0.11287 \\
    N     & 0.00000 & 0.00000 & 0.00000 & 0.00000 & 0.00000 & 0.00000 & 0.00000 & 0.00000 & 0.00000 \\
    H\footnotemark[1]    & 0.00000 & 0.00000 & 1.02056 & 0.00000 & 0.00000 & 0.78922 & 0.00000 & 0.00000 & 1.05819 \\

    \end{tabular}%
    \end{ruledtabular}
    \footnotetext[1]{This is the center of the muonic basis set for the TC-DFT optimized geometries. }
  \label{tab:s23}%
\end{table*}%

\bibliography{}